\documentclass[aip,apl,reprint,floatfix]{revtex4-1}
\usepackage{graphicx}
\usepackage[version=3]{mhchem} % Formula subscripts using \ce{}
\draft % marks overfull lines with a black rule on the right
\begin{document}

% Use the \preprint command to place your local institutional report number 
% on the title page in preprint mode.
% Multiple \preprint commands are allowed.
%\preprint{}

\title{AC Response Across the Metal–Insulator Transition of YBCO Josephson Junctions Fabricated with a Helium-Ion Beam} %Title of paper

% repeat the \author .. \affiliation  etc. as needed
% \email, \thanks, \homepage, \altaffiliation all apply to the current author.
% Explanatory text should go in the []'s, 
% actual e-mail address or url should go in the {}'s for \email and \homepage.
% Please use the appropriate macro for the type of information

% \affiliation command applies to all authors since the last \affiliation command. 
% The \affiliation command should follow the other information.

\author{Adhilsha Parachikunnumal}
\author{Nirjhar Sarkar}
\author{Aravind Rajeev Sreeja}
\author{Sreekar Vattipalli}
\author{Rochelle Qu}
\author{Jay~C.~LeFebvre}
\author{Roger K. Lake}
\author{Shane A. Cybart}
\email{cybart@ucr.edu}
\affiliation{Department of Electrical and Computer Engineering, University of California Riverside}

%\author{}
%\email[]{Your e-mail address}
%\homepage[]{Your web page}
%\thanks{}
%\altaffiliation{}
%\affiliation{}

% Collaboration name, if desired (requires use of superscriptaddress option in \documentclass). 
% \noaffiliation is required (may also be used with the \author command).
%\collaboration{}
%\noaffiliation

\date{\today}

\begin{abstract}
Using focused helium ion beam (FHIB) irradiation, we fabricated 
in-plane, high-$T_c$ 
YBa$_2$Cu$_3$O$_{7-\delta}$ (YBCO) Josephson junctions.
By varying the dose of the irradiation, we tune the junction barriers from 
metallic (SNS) to insulating (SIS) and investigate how this transition 
affects microwave-driven dynamics.
As the barrier transitions from metallic to insulating, the oscillatory response 
of the Shapiro steps to 
the RF power changes dramatically. 
On either side of the metal-insulator transition, the devices exhibit clean 
integer Shapiro steps without half-integer features, 
demonstrating that the 
current--phase relation is dominated by the first harmonic and that the
excess  current is  minimal. 
The current-voltage response is well-described by the resistively, capacitively shunted junction model assuming a single-harmonic current--phase relation.
This behavior indicates well-controlled junction properties suitable for a wide 
range of superconducting electronics, including detectors, mixers, and high-density integrated circuits.
\end{abstract}

\pacs{}% insert suggested PACS numbers in braces on next line

\maketitle %\maketitle must follow title, authors, abstract and \pacs

% Body of paper goes here. Use proper sectioning commands. 
% References should be done using the \cite, \ref, and \label commands
%\section{}
%\label{}
%\subsection{}
%\subsubsection{}
% (high-$T_\text{C}$)
%$^+$
% Introduction (you can delete this_i just wrote it in the beginging itself )
%\subsection{{$Introduction_{\textit{will be removed after corrections}}$}}

The AC Josephson effect provides a direct link between frequency and voltage, forming the foundation for primary voltage standards used worldwide by national metrology institutes.~\cite{Hamilton2000JosephsonVoltageStandards} These standards rely on the generation of precise, quantized voltages, known as Shapiro steps,~\cite{shapiro1963} under microwave irradiation. Beyond metrology, the AC response of Josephson junctions is essential for a variety of superconducting RF and microwave devices, including filters,\cite{filter-sam} mixers,\cite{mixer-Du-csiro} and detectors.\cite{HTcJJde}

In many high-transition temperature ($T_c$) Josephson junctions, particularly weak-link type devices, low impedance and strong Andreev reflections give rise to excess current,\cite{BTK} and higher harmonic components in the current--phase relation (CPR).~\cite{ half-integ-theory, hilgenkamp} These effects often manifest as half-integer Shapiro steps and non-ideal AC behavior. While numerous high-$T_c$ junction designs have been studied, 
\cite{hilgenkamp, point_contact,du2014fabrication, cybart2005josephson} controlling and understanding the AC response across different barrier types remains a central area of research.\cite{franch_scientificreport,Chinease-chen2022high}
Focused helium ion beam (FHIB) irradiation offers a powerful method to define Josephson junction barriers with nanometer-scale precision.\cite{cybart2015nano} By varying the ion dose, the barrier can be smoothly tuned from normal metal to insulator for comparison of different devices constructed from a single YBa$_2$Cu$_3$O$_{7-\delta}$ (YBCO) thin film.~\cite{cybart2015nano,cho2018superconducting,shaznjay,ethan_squid} Ion irradiated junctions have demonstrated excellent stability,\cite{cybartstability,germany-muller2019josephson} reproducibility,\cite{ cho2015yba2cu3o7, germany-muller2019josephson} and scalability,\cite{jay2021LargeScaleHIBL, Cybart2009nanolette, longjay} and they have been applied to a wide range of superconducting circuits, including SQUIDs,~\cite{ethan_squid, Cho20192DHighTcSQUIDArrays, haosquidarray-APL} digital high-$T_c$ electronics,~\cite{hanqfp, hna_SFQ} and quantum devices.\cite{VectorSubstrateJJ, YBCOJosephsonDiode_dieter, Max_HeFIBJJA} %amplifiers,\cite{haosquidarray-APL}

Previous studies have provided a foundational understanding of the AC response of FHIB junctions.~\cite{franch_scientificreport,Chinease-chen2022high,cortez2019tuning} Building on this important work, we extend the investigation to explore what happens as the barrier changes from metallic to insulating. By fabricating a series of FHIB-defined junctions spanning the metal-insulator-transition, we systematically study how the Shapiro step structure and RF power dependence evolve. This approach provides insight into how barrier properties govern AC Josephson dynamics and aids the future development of high-$T_c$ FHIB devices for microwave applications.

FHIB Josephson junctions for this work were fabricated from commercially grown YBCO films deposited on cerium oxide-buffered, $r$-plane, 50-mm sapphire wafers.\cite{YT_Multilayers} The YBCO layer was 40-nm thick, and before breaking vacuum, a 200-nm gold (Au) layer was deposited \textit{in-situ} to provide low-resistance contacts.

\begin{figure}[!b]         
\centering
\includegraphics[width=0.7\linewidth,keepaspectratio,trim={0cm 0cm 0cm 0cm},clip]{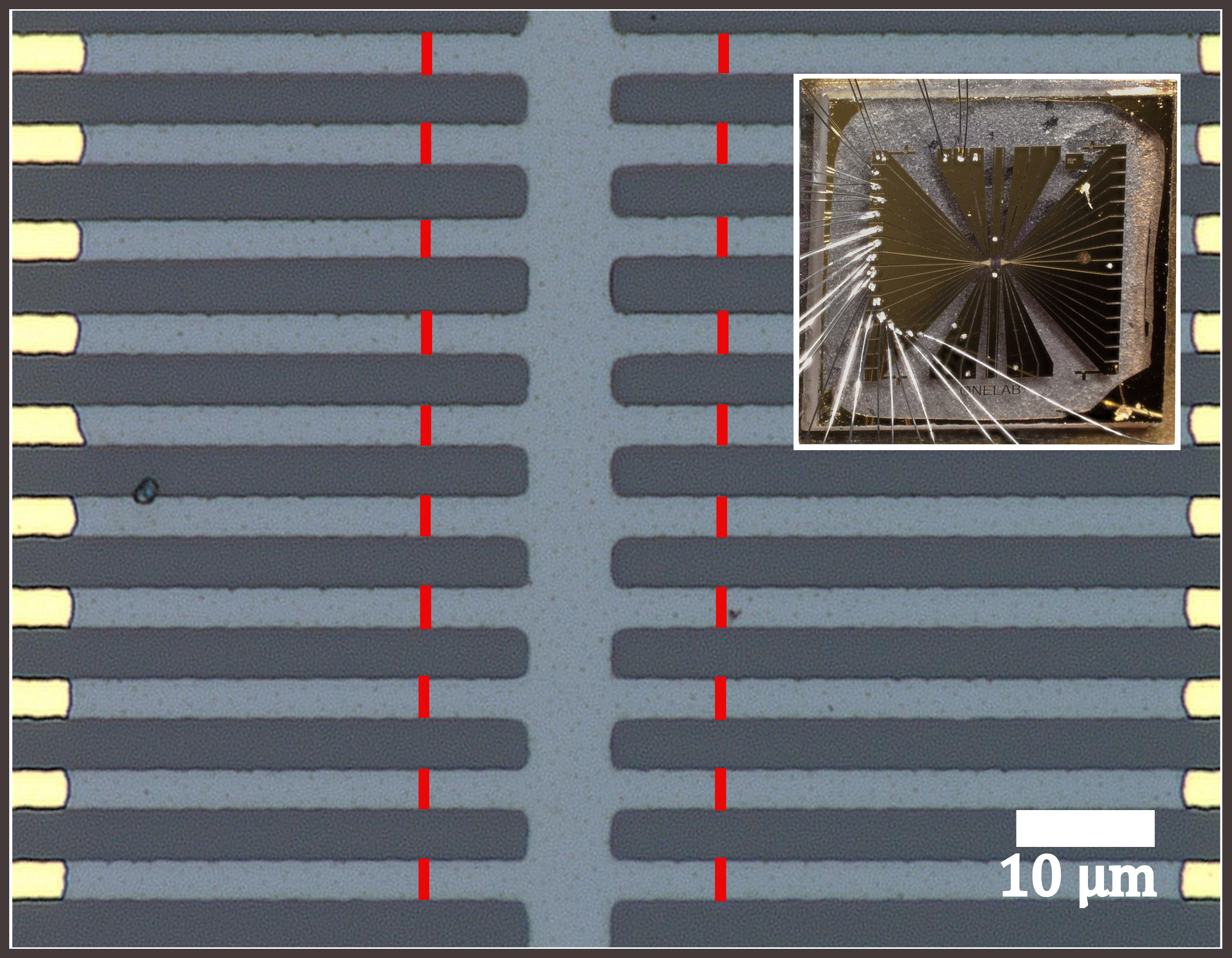}
\setlength\belowcaptionskip{0pt}

\caption{Microscope image of a 100~$\times$~100~$\mu$m region showing twenty 4-$\mu$m-wide YBCO bridges where the Au layer was selectively removed to expose the underlying YBCO prior to focused helium-ion-beam irradiation of the junctions (red lines). The inset shows the full 5~$\times$~5~mm chip containing the four-point bridge structure.}

\label{fig:1}

\end{figure}

The design of the test samples is shown in Fig.~\ref{fig:1}.  Wafers were diced into 5-mm~$\times$~5-mm chips (Fig.~\ref{fig:1} inset). Each chip was patterned using laser lithography, and both the gold and YBCO layers were etched together with a broad-beam argon ion mill to define 4-$\mu$m-wide bridges (Fig.~\ref{fig:1}).
%%%%%%%%%%%%%%%%%%%%%%%%%%%%%%%%%%%%%%%%%%%%%%%%%%%%%%%%%%%%%%
%%%%%%%%%%%%%%%%%%%%%%%%%%%%%%%%%%%%%%%%%%%%%%%%%%%%%%%%%%%%%%
%%%%%%%%%%%%%%%%%%%%%%%%%%%%%%%%%%%%%%%%%%%%%%%%%%%%%%%%%%%%%%
%%%%%%%%%%%%%%%%%%%%%%%%%%%%%%%%%%%%%%%%%%%%%%%%%%%%%%%%%%%%%%
To form junctions, the YBCO under the gold was uncovered at the intended site for FHIB irradiation. This was done using a second lithography step to selectively remove the gold in this area with a potassium iodide (KI$^+$) wet etch.

FHIB irradiation was performed at an accelerating voltage of 35~kV with a $\sim$1~nm beam spot. The beam was scanned across the 4-$\mu$m-wide YBCO bridges to define narrow irradiated barriers; prior work established the barrier width to be $<3$~nm.\cite{yan-tin_jj_width} Devices written at doses of 1.6 and 2.0~$\times$10$^{16}$~ions/cm$^2$ exhibited superconductor--normal--superconductor (SNS) behavior, whereas devices written at 2.8, 3.0, and 3.4~$\times$10$^{16}$~ions/cm$^2$ exhibited superconductor--insulator--superconductor (SIS) behavior. A device created with 2.4~$\times$10$^{16}$~ions/cm$^2$ displayed intermediate (crossover) characteristics.

%%%%%%%%%%%%%%%%%%%%%%%%%%%%%%%%%%%%%%%%%%%%%%%%%%%%%%%%%%%%%%
%%%%%%%%%%%%%%%%%%%%%%%%%%%%%%%%%%%%%%%%%%%%%%%%%%%%%%%%%%%%%%
\begin{figure}[!t]          % one-column float; !t usually places it at top of the column
\centering
\includegraphics[width=1\linewidth,keepaspectratio,trim={0cm 0cm 0cm 0cm},clip]{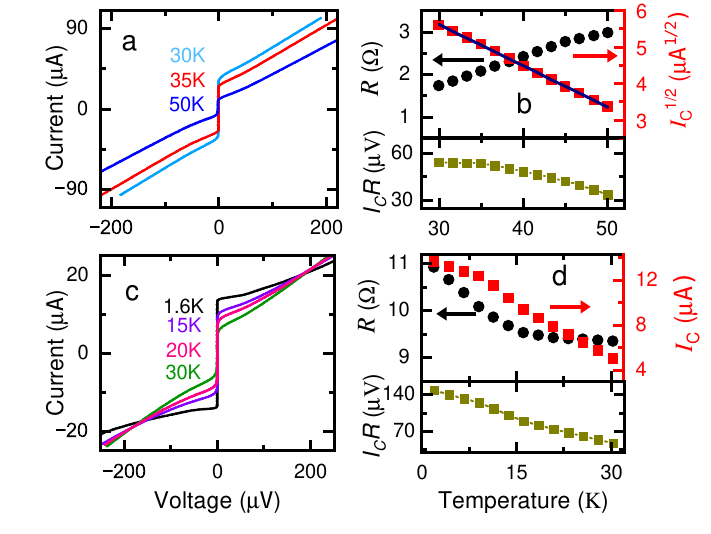} 
\setlength\belowcaptionskip{-10pt}
   \setlength\abovecaptionskip{-10pt}
\caption{ Transport measurements of FHIB-fabricated Josephson junctions illustrating metallic and insulating barrier characteristics. (a) \textit{I--V} characteristics of a SNS junction measured at 30, 35, and 50~K created with  2.0~$\times$10$^{16}$ ions/cm$^2$. (b) $I_C^{1/2}$ , $R$ and their product vs temperature for the SNS junction. (c) \textit{I--V} characteristics of  a SIS junction measured at 1.6, 15, 20, and 30~K created with 3.0~$\times$10$^{16}$ ions/cm$^2$. (d) $I_C$, $R$ and their product vs temperature for the SIS junction.}
\label{IV_R}

\end{figure}
%%%%%%%%%%%%%%%%%%%%%%%%%%%%%%%%%%%%%%%%%%%%%%%%%%%%%%%%%%%%%%
%%%%%%%%%%%%%%%%%%%%%%%%%%%%%%%%%%%%%%%%%%%%%%%%%%%%%%%%%%%%%%

Four-point measurements were performed with the junctions cooled in a Quantum Design Dyna-Cool PPMS equipped with a $\mu$-metal magnetic shield to reduce flux trapping. $\textit{I--V}$ characteristics were measured by current biasing the sample with a custom-built differential line driver; device signals were then amplified using SR560 battery-powered low-noise preamplifiers, and the amplified signals were recorded with a National Instruments data-acquisition system.

Figure~\ref{IV_R} compares transport in a metallic-barrier (SNS) junction and an insulating-barrier (SIS) junction. Figures~\ref{IV_R}(a) and \ref{IV_R}(c) show representative \textit{I--V} curves at (30, 35, and 50~K) and (1.6, 15, 20, and 30~K) for the SNS and SIS junctions respectively.  Additional temperatures were measured but omitted from the graph for clarity. Each \textit{I--V} curve was fit with the resistively and capacitively shunted junction (RCSJ) model to extract the critical current ($I_C$) and resistance ($R$). The corresponding $I_C R$  product is shown below, which is plotted versus temperature in Figs.~\ref{IV_R}(b) and \ref{IV_R}(d) for all of the SNS and SIS devices, respectively. 

For the SNS device, we plot ($I_C^{1/2}$) to highlight that the $I_C(T)$ follows a strongly coupled weak-link trend consistent with the de~Gennes~\cite{deGennes1964} description of SNS junctions, $I_C(T) = I_{C0}\left(1 - (T /\ T_c)\right)^2$.
The $R(T)$ in Fig.~\ref{IV_R}(b) decreases approximately linearly with decreasing temperature over the measured range, as typical for metallic YBCO.\cite{cybart2015nano} In contrast, the SIS device (Fig.~\ref{IV_R}(d)) shows $R(T)$ increasing as temperature decreases and $I_C(T)$ approaching a low-temperature plateau, consistent with YBCO SIS junctions.\cite{cybart2015nano} The distinct $I_C(T)$ and $R(T)$ trends across these two devices, together with intermediate-dose devices discussed below, demonstrate that focused helium-ion irradiation tunes the barrier continuously through the metal--insulator transition.

To investigate the AC Josephson response, microwave signals at 18~GHz were introduced to the junctions using a single conductor open coax cable for an antenna positioned next to the device inside the PPMS. The RF power was controlled by applying a 1~mHz ramp signal to the external amplitude modulation input of the waveform synthesizer. The devices were measured using the same four-point configuration described previously, with \textit{I--V} characteristics shown in Fig.~2.

\begin{figure}[ht]
    \centering
    \includegraphics[
        width=1.0\linewidth,
        keepaspectratio,
        trim=0pt 0pt 0pt 0pt,
        clip
    ]{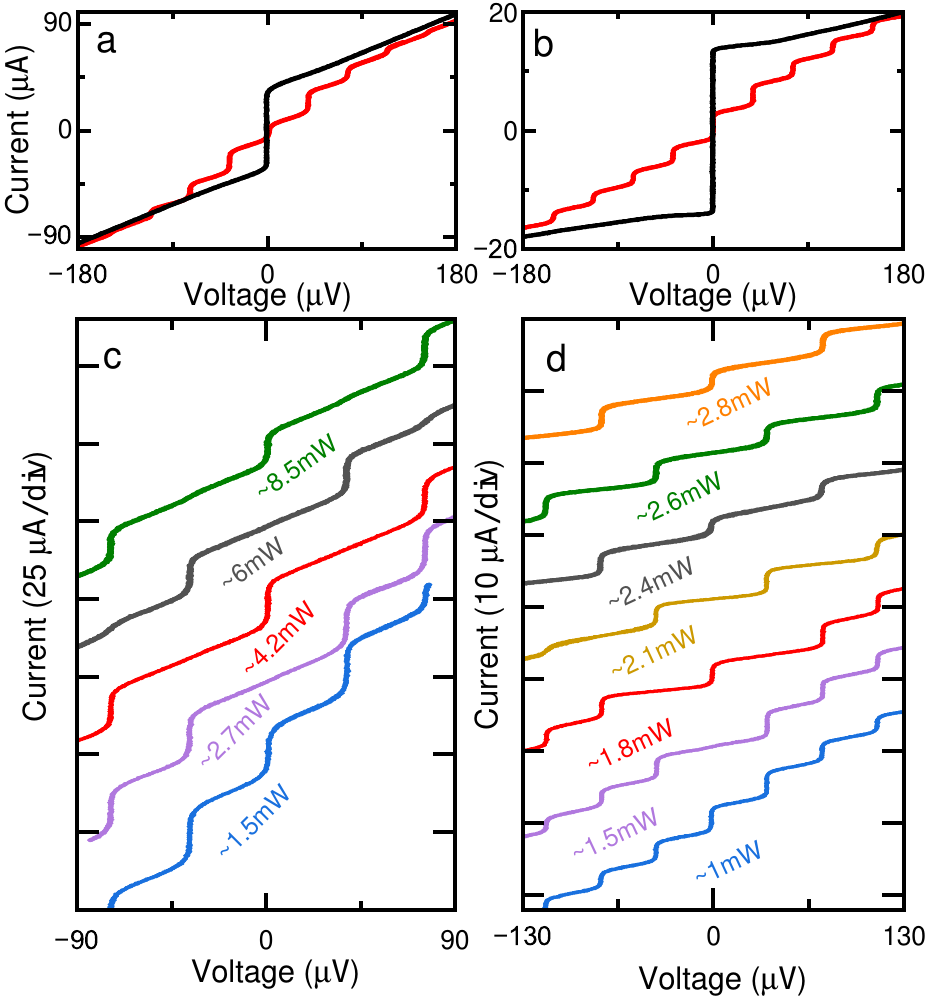}
    \setlength\belowcaptionskip{-10pt}        \setlength\abovecaptionskip{-10pt}
    % Adjust width as needed
\caption{\textit{I--V }characteristics of an SNS (a) at 30~K and an SIS (b) at 1.6~K junctions, measured with 18~GHz irradiation (red), and without (black). (c) and (d) show the \textit{I--V} characteristics of the same SNS  (30~K) and SIS (1.6~K) junctions, respectively, for different applied microwave powers. These data illustrate that the step heights go to zero at the minima eg. n=1 at 4.2~mW and 8.5~mW for the SNS junction. }
%Color plot illustrating the differential resistance ($dV/dI$) as a function of voltage and source power for metallic and insulating junctions. f) The waveform resulting from Shapiro step oscillations reflects the response to variations in microwave source power from step 0 to step 4 for  insulating junctions. }
\label{fig:fig3}
\end{figure}

Figure~\ref{fig:fig3} compares the microwave response of a representative SNS and SIS junction. Figures~\ref{fig:fig3}(a) and \ref{fig:fig3}(b) show \textit{I--V} characteristics measured with an 18~GHz RF signal (red) and without (black) for the SNS (30~K) and SIS (1.6~K) junctions respectively. In both junctions, we observe well-defined integer Shapiro steps and no half-integer steps or other parasitic features. 
In Figs.~\ref{fig:fig3}(c) and ~\ref{fig:fig3}(d) we show the \textit{I--V} characteristics at the same temperatures of the SNS and SIS junctions, respectively for different applied microwave powers. The curves are shifted vertically for clarity.  The complete extinction of the step amplitudes is observed at each oscillation minimum, dropping cleanly to zero without any residual structure. These data provide a clear hallmark of high-quality Josephson behavior. The complete evolution of the RF response is shown in the differential resistance maps of Fig.~(S1) in the 
Supplementary Materials (SM).

%The complete evolution of the RF response is shown in the differential resistance maps of Fig.~\ref{color plot}. These plots display $dV/dI$ as a function of voltage and microwave power, revealing the Shapiro steps as sharp, uniform bands. The maps provide a compact visualization of the junction dynamics, confirming the absence of half-integer steps and other spurious resonances. This type of presentation highlights the clean, ideal AC Josephson behavior of both junction types and underscores the high quality of the FHIB-fabricated devices.
%%%%%%%%%%%%%%%%%%%%%%%%%%%%%%%%%%%%%%%%%%%%%%%%%%%%%%%%%%%%%%
%%%%%%%%%%%%%%%%%%%%%%%%%%%%%%%%%%%%%%%%%%%%%%%%%%%%%%%%%%%%%%
%%%%%%%%%%%%%%%%%%%%%%%%%%%%%%%%%%%%%%%%%%%%%%%%%%%%%%%%%%%%%%
%We note that the Shapiro step heights became zero at certain microwave, and no half-integer steps are observed. 
%We attribute the reduced step height primarily to weak microwave coupling, which yields a small effective rf voltage at the junction.~\cite{hamilton1970}
%%%%%%%%%%%%%%%%%%%%%%%%%%%%%%%%%%%%%%%%%%%%%%%%%%%%%%%%%%%%%%

%%%%%%%%%%%%%%%%%%%%%%%%%%%%%%%%%%%%%%%%%%%%%%%%%%%%%%%%%%%%%%

\begin{figure}[ht]
    \centering
    \includegraphics[width=1.0\linewidth,keepaspectratio]{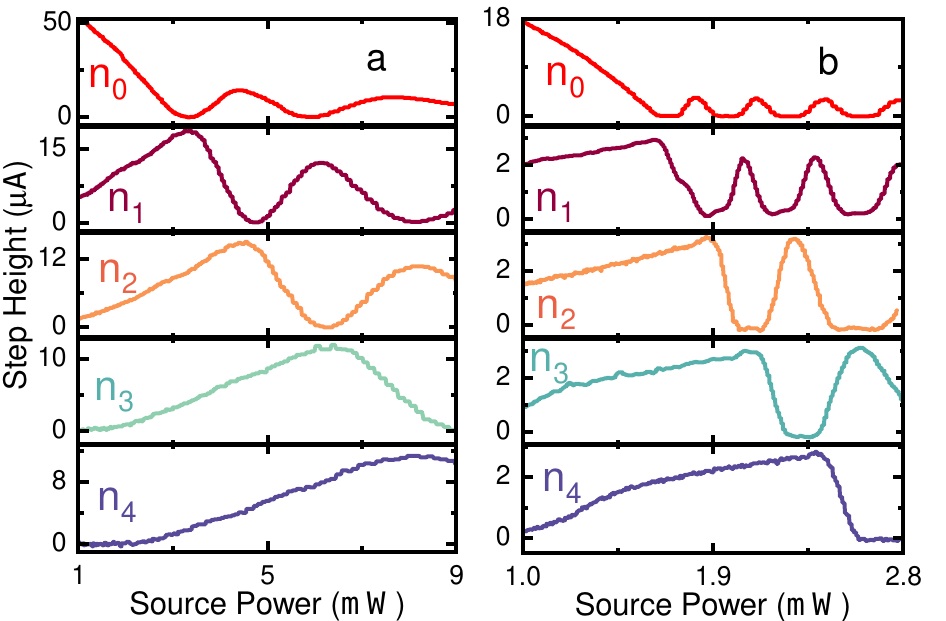}
    \setlength\belowcaptionskip{0pt}% Adjust width as needed
\caption{Measured step heights of the zeroth through fourth order Shapiro steps as a function of applied microwave source power for the SNS junction (a) with $I_CR$ 54~$\mu V$ at 30~K and the SIS junction (b) with $I_CR$ 150~$\mu V$ at 1.6~K, showing Bessel-like oscillations.}
%Color plot illustrating the differential resistance ($dV/dI$) as a function of voltage and source power for metallic and insulating junctions. f) The waveform resulting from Shapiro step oscillations reflects the response to variations in microwave source power from step 0 to step 4 for  insulating junctions. }
\label{bessel}
\end{figure}

Figures~\ref{bessel}(a) and \ref{bessel}(b) plot the measured step heights ($n = 0$ through $n = 4$) as a function of microwave source power for the SNS and SIS junctions respectively. The step heights follow the expected Bessel-like oscillations of the AC Josephson effect and are clearly resolved up to the fourth order. 
%A key distinction between the two junction types is that SNS devices require roughly five times more applied power to traverse a full oscillation cycle compared to SIS devices. This reflects fundamental differences in their barrier properties and energy scales: metallic junctions are harder to drive into full oscillation, while insulating junctions respond at much lower power levels.

The Kautz model provides a framework for rf-driven Josephson junctions operated in the low reduced-frequency regime $\Omega = f/f_c < 1$. Using this model, we compare our experimentally extracted Shapiro step currents with the predicted ideal values. For the SNS and SIS junctions, the measured $I_cR$ products are shown in Figs.~\ref{IV_R}(b) and \ref{IV_R}(d), corresponding to characteristic frequencies $f_c \approx 26~\mathrm{GHz}$ and $f_c \approx 72~\mathrm{GHz}$, respectively. At the measurement frequency $f = 18~\mathrm{GHz}$, this yields $\Omega \approx 0.69$ (SNS) and $\Omega \approx 0.25$ (SIS). Using a thin-film in-plane estimate for the Josephson penetration depth,\cite{Tolpygo1996CriticalCurrents} we obtain $\lambda_J \approx 5~\mu\mathrm{m}$ for the SNS junction and $\lambda_J \approx 8.7~\mu\mathrm{m}$ for the SIS junction. From Figs.~\ref{bessel}(a) and \ref{bessel}(b), we extract the step heights of the $n=1$ ($\Delta I_1$) and $n=2$ ($\Delta I_2$) Shapiro steps and obtain simultaneous maximum step currents $\Delta I_{\max} \approx 12~\mu\mathrm{A}$ for the SNS junction at $P_{\mathrm{RF,SNS}} \approx 3.77~\mathrm{mW}$ and $\Delta I_{\max} \approx 2.4~\mu\mathrm{A}$ for the SIS junction at $P_{\mathrm{RF,SIS}} \approx 1.5~\mathrm{mW}$. Comparing these values to the Kautz point-contact prediction,\cite{kautz1995shapiro} $\Delta I_{\max}^{\mathrm{Kautz}} \approx 32.4~\mu\mathrm{A}$ (SNS) and $5.8~\mu\mathrm{A}$ (SIS), we obtain percentage differences of $\sim 62\%$ and $\sim 59\%$, respectively. Since the ion-irradiated junction length is on the 3-nm scale (short in the current-flow direction)\cite{yan-tin_jj_width} and the lateral width is comparable to $\lambda_J$, the junctions lie near the crossover between the point-contact and long junction limits. We therefore also compare our data to the corresponding long junction Kautz reference,\cite{kautz1995shapiro} for which $\Delta I_{\max}^{\mathrm{Kautz}} \approx 28.4~\mu\mathrm{A}$ (SNS) and $\approx 2.8~\mu\mathrm{A}$ (SIS), giving percentage differences of $\sim 58\%$ (SNS) and $\sim 14\%$ (SIS). Overall, these percentage differences likely reflect a combination of thermal noise in our measurements and the in-plane planar junction geometry, which falls between the ideal point-contact and strongly distributed limits; a dedicated model incorporating thin-film in-plane electrodynamics and rf coupling is therefore required for a fully quantitative description.

\begin{figure}[!t]   % <-- H means "place it exactly here"
\centering
\includegraphics[width=1.0\linewidth,keepaspectratio]{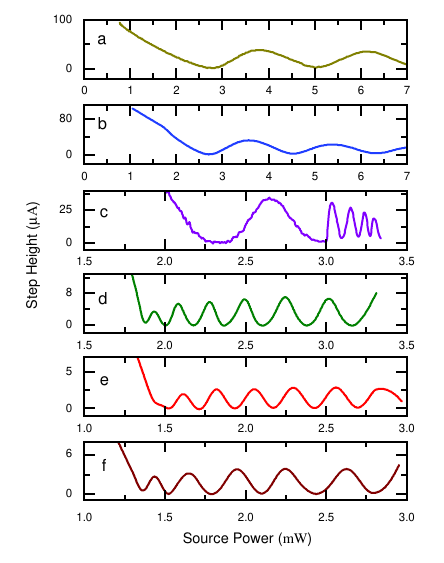}
\setlength\belowcaptionskip{-10pt}
   \setlength\abovecaptionskip{-15pt}
\caption{ Shapiro step height versus microwave source power for junctions on  the same chip written with doses 1.6 (a) and, 2.0 (b) are measured at 60~K, 2.4 (c) at 30~K, and 2.8 (d), 3.0 (e), and 3.4~$\times$10$^{16}$ ions/cm$^2$ (f) all  measured at 2~K .}
%Temperature dependence of the microwave oscillation in a SIS junction from 40 K of 7.3 $\Omega$  to 1.6 K of 17 $\Omega$ , it shows how the amplitude and the period of the  wavelet  changes with temperature and resistance.}
\label{fig6}
\end{figure}

To analyze how the oscillation periodicity of the Shapiro step depends on source power, we fabricated another device with doses of $1.6$, $2.0$, $2.4$, $2.8$, $3.0$, and $3.4 \times 10^{16}$~ions/cm$^2$. Figure~\ref{fig6} displays the  Shapiro step height as a function of source power for the zeroth-step oscillation in junctions spanning a wide range of resistances.  SNS junctions fabricated with lower doses of $1.6$ (Fig.~\ref{fig6}(a)) and $2.0 \times 10^{16}$~ions/cm$^2$ (Fig.~\ref{fig6}(b)) show resistances at $60$~K of approximately $0.5~\Omega$ and $0.8~\Omega$, and SIS junctions made with doses of $2.8$ (Fig.~\ref{fig6}(d)), $3.0$ (Fig.~\ref{fig6}(e)), and $3.4 \times 10^{16}$~ions/cm$^2$ (Fig.~\ref{fig6}(f)) have resistances at $2$~K of approximately $6~\Omega$, $10~\Omega$, and $12~\Omega$, respectively. A crossover junction irradiated with $2.4 \times 10^{16}$~ions/cm$^2$ (Fig.~\ref{fig6}(c)) has a resistance of $R \approx 2~\Omega$ at $30$~K and occupies an intermediate regime in which the ion-irradiated region transitions from metallic to insulating character. This $2.4 \times 10^{16}$~ions/cm$^2$ device is the only junction that exhibits a clear crossover oscillation pattern: at low microwave power, it behaves like an SNS junction, showing the long-period oscillations characteristic of a metallic barrier, whereas at higher powers, as the junction is driven to larger voltages, the oscillations abruptly switch to the short-period behavior typical of an insulating (SIS) junction, as shown in the inset in Fig.~\ref{fig6}(c). We have reproduced these behaviors across three chiplets with same ion irradiation; although the absolute resistance values vary slightly from chip to chip, the oscillation periodicity, the dose-dependent evolution, and in particular the $2.4 \times 10^{16}$~ions/cm$^2$ crossover signature are robust and reproducible. %The normalized power spacing of the zeroth Shapiro step therefore provides a sensitive microwave probe of the ion-irradiated barrier, revealing a systematic evolution from SNS to SIS behavior and identifying $2.4 \times 10^{16}$~ions/cm$^2$ as a well-defined crossover dose at the metal--insulator transition.

%%%%%%%%%%%%%%%%%%%%%%%%%%%%%%%%%%%%%%%%%%%%%%%%%%%%%%%%%%%%%%
%%%%%%%%%%%%%%%%%%%%%%%%%%%%%%%%%%%%%%%%%%%%%%%%%%%%%%%%%%%%%%

%%%%%%%%%%%%%%%%%%%%%%%%%%%%%%%%%%%%%%%%%%%%%%%%%%%%%%%%%%%%%%
%%%%%%%%%%%%%%%%%%%%%%%%%%%%%%%%%%%%%%%%%%%%%%%%%%%%%%%%%%%%%%
\begin{figure}[!t]   % <-- H means "place it exactly here"
\centering
\includegraphics[width=1.0\linewidth,keepaspectratio]{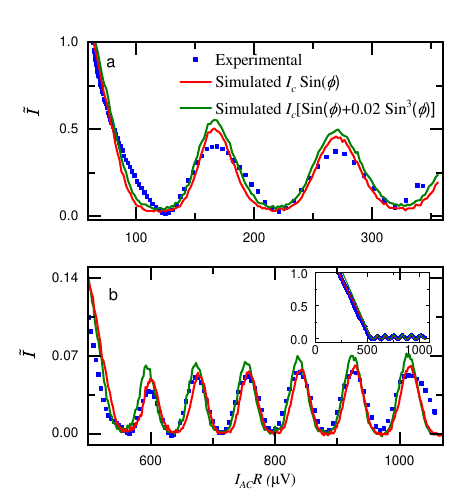}
\setlength\belowcaptionskip{-10pt}
   \setlength\abovecaptionskip{-10pt}
 \caption{Comparison of experimental data(blue scatter) and RCSJ model calculations for an SNS junction (a) from Figs~\ref{fig6}(a) and an SIS junction (b) from Fig~\ref{fig6}(e). Both datasets are fit using an ideal single-harmonic sinusoidal CPR (solid red lines) and a  CPR with a third harmonic (dashed green lines). Experimental data are plotted (Fig.~\ref{fig6}) versus source power rescaled by a constant factor, mapping the effective RF power $P \propto I_\mathrm{AC}^2 R$ onto the effective voltage-drive $i_\mathrm{ac} = I_\mathrm{AC} R$ for direct comparison with theory. The vertical axis has been  normalized to $\tilde{I}$ ($I_\mathrm{step}/I_{\mathrm{step, max~at~lowest~I_{\mathrm{AC}}}R}$).The inset shows full range of the experimental and simulated data of SIS junction}
% and a non-sinusoidal CPR (dashed green lines)
%Temperature dependence of the microwave oscillation in a SIS junction from 40 K of 7.3 $\Omega$  to 1.6 K of 17 $\Omega$ , it shows how the amplitude and the period of the  wavelet  changes with temperature and resistance.}
\label{fig7}
\end{figure}
%%%%%%%%%%%%%%%%%%%%%%%%%%%%%%%%%%%%%%%%%%%%%%%%%%%%%%%%%%%%%%

%%%%%%%%%%%%%%%%%%%%%%%%%%%%%%%%%%%%%%%%%%%%%%%%%%%%%%%%%%%%%%

%Paragraph 1 and or 1-2
%Introduction : To test for anharmonicty we have performed numeraicla simulations.
    %Solve a modifieds RSCJ model
    %noise term
    %Rscj equation
    % allows for higher harminic content
    % other model details--how you solve it , eg solver python 
%concluding sentence : some take away sentence

%Pargraph 2 or 3
% introduction : Figure 7 shows .....I vs Iac
    %your model vs the data. details blue squares symbols model solid lines. how many points , fitting ranges and why is it ok to show as a solid line. 
        %how did you modify the experimentla data
    %7a SNS her explain the two traces ...what the cirteria used to chose the coffecitne. 1. is pure sin phi....0.02...error ananlysi.
    %7b shows th eSIS  and th einset shows the full range. .. we couyldnty find a third harmionic solution thats was better but here is the same one from 7a to show you what happens
% conclusion there are no other harmincs and it supoiorts the expermeitel data that say there are no othe rharmionics. 

To test for possible anharmonicity in the current–phase relation, we perform numerical 
simulations of the Shapiro step amplitudes as a function of the ac drive amplitude. 
Because the experiment is current biased and the main observable is the Shapiro step 
amplitude extracted from dc current–voltage characteristics under RF irradiation, 
we use a 
modified resistively and capacitively shunted junction (RCSJ) \cite{mccumber1968effect} 
model with thermal noise, which directly provides current–voltage curves for a current-
biased Josephson junction. The model details are described in the supplementary materials.

The junction parameters 
used in the simulations are chosen to match the experimental devices: for the SNS 
junctions we use $C = 10^{-16}\,~\mathrm{F}$, $I_c=170~\mu A$ and $R = 0.4\,~\Omega$, while 
for the SIS junctions \cite{yan-tin_jj_width} we use $C = 2.65\times10^{-15}\,~\mathrm{F}$, $I_c=70~\mu A$ and $R = 8.8\,
%%%Add Reference YC
~\Omega$. From the resulting $I$–$V$ characteristics we extract the Shapiro step amplitudes 
as a function of $i_{\mathrm{ac}}$, with particular attention to the zero-voltage ($n=0$) 
step.

Figure~\ref{fig7} summarizes the comparison between experiment and the stochastic RCSJ 
model by showing the normalized zero-voltage ($n=0$) Shapiro step amplitude $\tilde{I}$ as 
a function of the ac drive. 
The blue squares represent the experimental data, and the 
solid curves are obtained from numerical solutions of Eq.~(S2) 
 in SM. 
Each theoretical curve consists of several hundred simulation points 
that are connected  by lines as a guide to the eye. 
Because the RF signal experiences attenuation and 
impedance mismatch between the microwave line and the junction, the nominal source power 
does not coincide with the power actually delivered to the device. 
To place experiment and 
theory on a common scale, we assume that the RF power at the junction is given by the 
measured source power divided by a single constant factor for each device, which 
effectively accounts for these losses. This defines an effective RF power in the model, 
which we convert to an ac current amplitude using $P \propto I_{\mathrm{AC}}^{2}R$ and 
then to the effective voltage-drive $I_{\mathrm{AC}}R$. The constant factor is calculated to be \(1.7\times 10^{-5}\) for the metallic junction and \(2.9\times 10^{-5}\) for the insulator junction. 
% The constants are very small because the $I_c^2R$ values are in the nW range, whereas the applied RF power is in the mW range.
The experimental points are therefore plotted as a function of source power rescaled by this 
constant factor, so that the horizontal axis can be compared directly with the theoretical 
dependence on $i_{\mathrm{ac}}$ and vertical axis has been  normalized to $\tilde{I} (I_\mathrm{step}/I_{\mathrm{step, max~at~lowest~I_{\mathrm{AC}}R}}$).

In Fig.~\ref{fig7}(a), the results for the SNS junction are displayed with two 
theoretical curves: one with a purely sinusoidal current–phase relation and one including 
a third-harmonic component with an amplitude equal to $2\%$ of the first harmonic. The $2\%
$ value is obtained by minimizing the discrepancy between the calculated Shapiro step 
amplitudes and the experimental data in the oscillatory region of $\tilde{I}
(i_{\mathrm{ac}})$. Fits that attempted to include the low-amplitude, nearly linear regime 
tended to overfit that part of the data while underfitting the oscillatory 
behavior, so only the oscillatory region was used to evaluate the fitting error. 
Within  this range, the third-harmonic term constitutes only a minor correction 
and does not 
appreciably improve the agreement compared with the purely sinusoidal model. This 
indicates that the SNS current–phase relation can be treated, to good approximation, as 
single harmonic.

Figure~\ref{fig7}(b) presents the corresponding analysis for the SIS junction, where  a 
purely sinusoidal current–phase relation reproduces the measured step amplitudes over the 
full range of rescaled drive. 
Fig.~\ref{fig7}(b) focuses on the high-drive region, 
while the inset shows the full drive range. For comparison, we also plot a curve with a 
$2\%$ third-harmonic component, identical to that used for the SNS junction. This clearly 
deviates from the data and demonstrates that adding a third harmonic does not improve the 
fit. 
Overall, Fig.~\ref{fig7} shows that both SNS and SIS junctions are well described by 
an almost purely sinusoidal current–phase relation, with only a negligible   ($\sim 2\%$) 
third-harmonic correction in the SNS case and no detectable higher-order term in the SIS 
junction. This observation is consistent with the clean integer Shapiro steps in 
Figs.~\ref{fig:fig3} and~(S1), confirming that both junctions display an 
almost ideal current phase relationship. 
%%%%%%%%%%

To account for the shorter apparent oscillation period of the SIS junction 
compared to the SNS junction in Fig.~\ref{fig6}, we compare the datasets using 
the RF voltage scale \(I_{\mathrm{AC}}R\). 
When the data are expressed in terms 
of this junction-relevant drive parameter (Fig.~\ref{fig7}), the apparent 
difference is strongly reduced and the oscillation periods become comparable. 
Quantitatively, the spacing between adjacent maxima is \(\Delta V \approx 101\,
~\mu\mathrm{V}\) for the SNS junction and \(\Delta V \approx 85\,~\mu\mathrm{V}\) 
for the SIS junction. 
This trend is consistent with a voltage-source description 
of the RF coupling: in this picture, the Josephson current in the presence of an 
applied RF voltage is given by Eq.~(S9), and the oscillation period 
is set by the ac voltage amplitude through the Bessel-function argument.

%%%%%%%%%
In summary,
using focused helium ion beam irradiation, we fabricated high-$T_c$ YBCO junctions that 
span the full range from metallic (SNS) to insulating (SIS) barriers. 
These devices 
exhibit clean integer Shapiro steps without half-integer features, demonstrating that the 
current--phase relation is dominated by the first harmonic and that excess current is 
minimal. 
This behavior indicates well-controlled junction properties suitable for a wide 
range of superconducting electronics, including detectors, mixers, and high-density 
integrated circuits.

\begin{acknowledgments}
This work was supported by the Air Force Office of Scientific Research under Grants FA9550-20-1-144, FA9550-23-1-0369 and the Department of Energy NNSA grant DE-NA0004106.
\end{acknowledgments}

% Create the reference section using BibTeX:
%\bibliography{bib}
%%%%%%%%%%%%%%%%%%%%%%%%%%%%%%%%%%%%%%%%%%%%%%%%%%%%%%%%
%merlin.mbs aipnum4-1.bst 2010-07-25 4.21a (PWD, AO, DPC) hacked
%Control: key (0)
%Control: author (8) initials jnrlst
%Control: editor formatted (1) identically to author
%Control: production of article title (-1) disabled
%Control: page (0) single
%Control: year (1) truncated
%Control: production of eprint (0) enabled
%

%%%%%%%%%%%%%%%%%%%%%%%%%%%%%%%%%%%%%%%%%%%%%%%%%%%%%%%%%%%%%%%%%%%%%%%%%%
\clearpage
%\documentclass[aip,apl,reprint]{revtex4-1}

%\usepackage{graphicx}
%\usepackage[version=3]{mhchem}
%\usepackage{hyperref} % needed because you define \theHfigure etc.

%\bibliographystyle{aipnum4-1}

%\begin{document}
% ---- Supplemental numbering: S1, S2, ... ----
%\clearpage
\setcounter{equation}{0}
\setcounter{figure}{0}
\setcounter{table}{0}
\setcounter{page}{1}

\renewcommand{\theequation}{S\arabic{equation}}
\renewcommand{\thefigure}{S\arabic{figure}}
\renewcommand{\thetable}{S\arabic{table}}

% If you use \label/\ref on figures/tables, this helps hyperref pick up the change:
\makeatletter
\makeatother

\begin{widetext}
\begin{center}
{\bf AC Response Across the Metal–Insulator Transition of YBCO Josephson Junctions Fabricated with a Helium-Ion Beam}\\ 
Adhilsha Parachikunnumal,
Nirjhar Sarkar,
Aravind Rajeev Sreeja,
Sreekar Vattipalli,
Rochelle Qu,
Jay~C.~LeFebvre,
Roger K. Lake,
Shane A. Cybart
(cybart@ucr.edu)\\
{Department of Electrical and Computer Engineering, University of California Riverside}
\end{center}
\end{widetext}

\section*{Supplementary Information}
%%%%%%%%%%%%%%%%%%%%%%%%%%%%%%%%%%%%%%%%%%%%%%%%%%%%%%%%%%%%%%
%%%%%%%%%%%%%%%%%%%%%%%%%%%%%%%%%%%%%%%%%%%%%%%%%%%%%%%%%%%%%%
\begin{figure}[ht]
    \centering
    \includegraphics[width=1\linewidth,keepaspectratio]{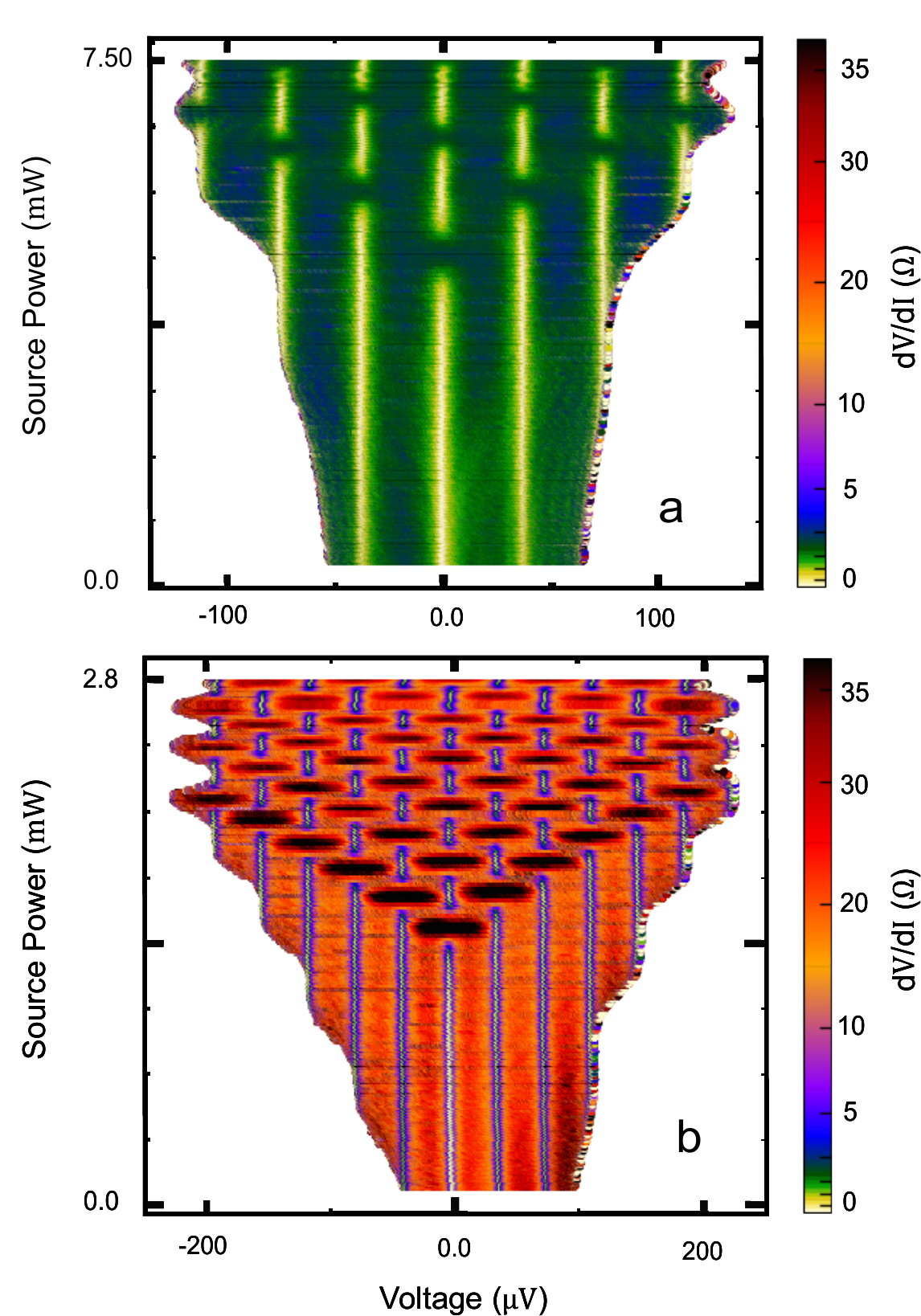}
    \setlength\belowcaptionskip{0pt}% Adjust width as needed
\caption{ Color maps of voltage vs microwave source power for (a) the SNS junction and (b) the SIS junction. These plots provide a detailed view of the microwave response, confirming the absence of half-integer steps and demonstrating the high quality of the junctions.}
%Color plot illustrating the differential resistance ($dV/dI$) as a function of voltage and source power for metallic and insulating junctions. f) The waveform resulting from Shapiro step oscillations reflects the response to variations in microwave source power from step 0 to step 4 for  insulating junctions. }
\label{color plot}
\end{figure}
%%%%%%%%%%%%%%%%%%%%%%%%%%%%%%%%%%%%%%%%%%%%%%%%%%%%%%%%%%%%%%
%%%%%%%%%%%%%%%%%%%%%%%%%%%%%%%%%%%%%%%%%%%%%%%%%%%%%%%%%%%%%%
The complete evolution of the RF response is shown in the differential resistance maps of Fig.~\ref{color plot}. These plots display $dV/dI$ as a function of voltage and microwave power, revealing the Shapiro steps as sharp, uniform bands. The maps provide a compact visualization of the junction dynamics, confirming the absence of half-integer steps and other spurious resonances. This type of presentation highlights the clean, ideal AC Josephson behavior of both junction types and underscores the high quality of the FHIB-fabricated devices.\\

\noindent
{\bf Resistively and Capacitively Shunted Junction (RCSJ) Model}

In this description, the Josephson element is shunted by a 
resistance $R$ and a capacitance $C$ connected in parallel, so that dissipation and 
capacitive effects are treated consistently. 
The applied RF field in the experiment is 
represented in the simulations by an ac current term $I_{\mathrm{AC}}\sin(\omega t)$ added 
to the dc bias $I_{\mathrm{DC}}$, and a possible third-harmonic contribution to the 
current–phase relation is included through the modified RCSJ equation
\begin{align}
\label{eq:RCSJ}
I_{\mathrm{DC}} + I_{\mathrm{AC}}\sin(\omega t)
&= I_{C}\big[\sin\phi(t) + \alpha_{3}\sin(3\phi(t))\big]\notag\\
&
+ \frac{\Phi_{0}}{2\pi R}\,\dot{\phi}(t)
+ \frac{\Phi_{0} C}{2\pi}\,\ddot{\phi}(t)
+ I_{N}(t),
\end{align}
where $\phi(t)$ is the superconducting phase difference, $I_{\mathrm{DC}}$ and 
$I_{\mathrm{AC}}$ are the DC and AC bias currents, $I_N(t)$ represents thermal noise $
\Phi_0$~is the flux quanta and $\alpha_{3}$ is the anharmonicity parameter of the current--
phase relation.
The characteristic frequency and Stewart–McCumber parameter are defined as $\omega_{c} = 
2\pi I_{C} R / \Phi_{0}$ and $\beta_{C} = 2\pi I_{C} R^{2} C / \Phi_{0}$, respectively. 
Using the normalized time $\tau = \omega_{c} t$, the normalized currents $i_{\mathrm{dc}} 
= I_{\mathrm{DC}}/I_{C}$ and $i_{\mathrm{ac}} = I_{\mathrm{AC}}/I_{C}$, and the normalized 
drive frequency $\Omega = \omega/\omega_{c}$, Eq.~\eqref{eq:RCSJ} can be written as
\begin{equation}
\begin{aligned}
\beta_C \ddot{\phi} + \dot{\phi}
+ \sin\textbf{}\!\left(\phi(\tau)\right)
+ \alpha_3 \sin\!\left(3\phi(\tau)\right)
= i_{\mathrm{dc}}&+ i_{\mathrm{ac}} \sin(\Omega\tau)\\
& + i_N(\tau),
\end{aligned}
\label{eq:rcsj_dimensionless}
\end{equation}
where dots now denote derivatives with respect to $\tau$. Thermal fluctuations of the shunt resistor are modeled as Johnson--Nyquist noise with
\begin{equation}
\langle I_{N}(t_{2}) I_{N}(t_{1}) \rangle = \frac{2k_{B}T}{R}\delta(t_{2}-t_{1})    
\end{equation}
 which in normalized units, with $i_{N}(\tau) = I_{N}(t)/I_{C}$ and $\tau = \omega_{c} t$, yields the correlation
\begin{equation}
    \label{eq:jn_noise_norm}
    \langle i_{N}(\tau_{2}) i_{N}(\tau_{1}) \rangle = 2D\,\delta(\tau_{2}-\tau_{1})
\end{equation}
where $D$ is a dimensionless noise strength.~\cite{aravind_sug} 
Equations~\eqref{eq:RCSJ}–\eqref{eq:jn_noise_norm} thus provide a complete 
stochastic RCSJ description of the junction, which we use to 
compute the Shapiro step amplitudes as a function of the ac drive amplitude.

To obtain the Shapiro step amplitudes from the stochastic RCSJ model, we solve 
Eq.~\ref{eq:rcsj_dimensionless} numerically. 
The equation is integrated using the Euler-Maruyama scheme for stochastic differential equations, with Gaussian noise consistent with 
the correlation in Eq.~\eqref{eq:jn_noise_norm}. 
For computational efficiency, the 
algorithm is implemented in Fortran, which allows us to perform long time traces and fine 
sweeps of the control parameters. 
For each value of the normalized ac amplitude 
$i_{\mathrm{ac}}$, we sweep the dc bias $i_{\mathrm{dc}}$ and integrate $\phi(\tau)$ over 
a long time interval. 
The normalized dc voltage is then obtained from the long-time 
average of the phase velocity,
\begin{equation}
\bar{v}(i_{\mathrm{dc}},i_{\mathrm{ac}}) = \big\langle \dot{\phi}(\tau) \big\rangle_{\tau},
\end{equation}
which corresponds to the physical voltage $V = I_{C}R\,\bar{v}$.\\

\noindent
{\bf Voltage-Source Model for an RF-Driven Josephson Junction}

While our devices are current driven, the voltage-source model
can be solved analytically and provides insight into the
Shapiro step oscillation frequency.
In this model, the Josephson device is driven by an ideal ac voltage source $V_s\cos\omega_s t$, so that
\begin{equation}
V(t)=V+V_s\cos\omega_s t.
\end{equation}
Combining the ideal Josephson current relation $i_j(t)=I_c\sin\phi(t)$ with the second Josephson relation $d\phi/dt=(2e/\hbar)V(t)$ gives
\begin{equation}
i_j(t)=I_c\sin\!\left[\int_{0}^{t}\frac{2e\,V(t')}{\hbar}\,dt' + \phi_0\right],
\end{equation}
where $\phi_0$ is a constant of integration. Substituting $V(t)$ and evaluating the integral yields
\begin{equation}
\label{eq:ic_voltage_source}
i_j(t)=I_c\sin\!\left[
\omega_J t+\left(\frac{2eV_s}{\hbar\omega_s}\right)\sin\omega_s t+\phi_0\right],
\end{equation}
with the Josephson oscillation frequency defined as $\omega_J=2eV/\hbar$.
Expanding out the sine function in terms of Bessel functions gives
\begin{equation}
i_j(t)=I_c\sum_{n=-\infty}^{\infty}(-1)^n\,J_n\!\left(\frac{2eV_S}
{\hbar\omega_S}\right)\sin\!\left[(\omega_j-n\omega_S)t+\phi_0\right],
\label{eq:Bessel_fn}
\end{equation}
Thus, the height of the Shapiro steps oscillate as a function of the
applied ac voltage $V_S$, which scales as $I_{AC}R$.

%\bibliography{bib}

\begin{thebibliography}{37}%
\makeatletter
\providecommand \@ifxundefined [1]{%
 \@ifx{#1\undefined}
}%
\providecommand \@ifnum [1]{%
 \ifnum #1\expandafter \@firstoftwo
 \else \expandafter \@secondoftwo
 \fi
}%
\providecommand \@ifx [1]{%
 \ifx #1\expandafter \@firstoftwo
 \else \expandafter \@secondoftwo
 \fi
}%
\providecommand \natexlab [1]{#1}%
\providecommand \enquote  [1]{``#1''}%
\providecommand \bibnamefont  [1]{#1}%
\providecommand \bibfnamefont [1]{#1}%
\providecommand \citenamefont [1]{#1}%
\providecommand \href@noop [0]{\@secondoftwo}%
\providecommand \href [0]{\begingroup \@sanitize@url \@href}%
\providecommand \@href[1]{\@@startlink{#1}\@@href}%
\providecommand \@@href[1]{\endgroup#1\@@endlink}%
\providecommand \@sanitize@url [0]{\catcode `\\12\catcode `\$12\catcode `\&12\catcode `\#12\catcode `\^12\catcode `\_12\catcode `\%12\relax}%
\providecommand \@@startlink[1]{}%
\providecommand \@@endlink[0]{}%
\providecommand \url  [0]{\begingroup\@sanitize@url \@url }%
\providecommand \@url [1]{\endgroup\@href {#1}{\urlprefix }}%
\providecommand \urlprefix  [0]{URL }%
\providecommand \Eprint [0]{\href }%
\providecommand \doibase [0]{http://dx.doi.org/}%
\providecommand \selectlanguage [0]{\@gobble}%
\providecommand \bibinfo  [0]{\@secondoftwo}%
\providecommand \bibfield  [0]{\@secondoftwo}%
\providecommand \translation [1]{[#1]}%
\providecommand \BibitemOpen [0]{}%
\providecommand \bibitemStop [0]{}%
\providecommand \bibitemNoStop [0]{.\EOS\space}%
\providecommand \EOS [0]{\spacefactor3000\relax}%
\providecommand \BibitemShut  [1]{\csname bibitem#1\endcsname}%
\let\auto@bib@innerbib\@empty
%</preamble>
\bibitem [{\citenamefont {Hamilton}(2000)}]{Hamilton2000JosephsonVoltageStandards}%
  \BibitemOpen
  \bibfield  {author} {\bibinfo {author} {\bibfnamefont {C.~A.}\ \bibnamefont {Hamilton}},\ }\href {\doibase 10.1063/1.1289507} {\bibfield  {journal} {\bibinfo  {journal} {Review of Scientific Instruments}\ }\textbf {\bibinfo {volume} {71}},\ \bibinfo {pages} {3611} (\bibinfo {year} {2000})}\BibitemShut {NoStop}%
\bibitem [{\citenamefont {Shapiro}(1963)}]{shapiro1963}%
  \BibitemOpen
  \bibfield  {author} {\bibinfo {author} {\bibfnamefont {S.}~\bibnamefont {Shapiro}},\ }\href {\doibase 10.1103/PhysRevLett.11.80} {\bibfield  {journal} {\bibinfo  {journal} {Physical Review Letters}\ }\textbf {\bibinfo {volume} {11}},\ \bibinfo {pages} {80} (\bibinfo {year} {1963})}\BibitemShut {NoStop}%
\bibitem [{\citenamefont {Watanabe}, \citenamefont {Dresselhaus},\ and\ \citenamefont {Benz}(2006)}]{filter-sam}%
  \BibitemOpen
  \bibfield  {author} {\bibinfo {author} {\bibfnamefont {M.}~\bibnamefont {Watanabe}}, \bibinfo {author} {\bibfnamefont {P.~D.}\ \bibnamefont {Dresselhaus}}, \ and\ \bibinfo {author} {\bibfnamefont {S.~P.}\ \bibnamefont {Benz}},\ }\href {\doibase 10.1109/TASC.2005.863533} {\bibfield  {journal} {\bibinfo  {journal} {IEEE Transactions on Applied Superconductivity}\ }\textbf {\bibinfo {volume} {16}},\ \bibinfo {pages} {49–53} (\bibinfo {year} {2006})}\BibitemShut {NoStop}%
\bibitem [{\citenamefont {Du}\ \emph {et~al.}(2012)\citenamefont {Du}, \citenamefont {Macfarlane}, \citenamefont {Pegrum}, \citenamefont {Zhang}, \citenamefont {Cai},\ and\ \citenamefont {Guo}}]{mixer-Du-csiro}%
  \BibitemOpen
  \bibfield  {author} {\bibinfo {author} {\bibfnamefont {J.}~\bibnamefont {Du}}, \bibinfo {author} {\bibfnamefont {J.~C.}\ \bibnamefont {Macfarlane}}, \bibinfo {author} {\bibfnamefont {C.~M.}\ \bibnamefont {Pegrum}}, \bibinfo {author} {\bibfnamefont {T.}~\bibnamefont {Zhang}}, \bibinfo {author} {\bibfnamefont {Y.}~\bibnamefont {Cai}}, \ and\ \bibinfo {author} {\bibfnamefont {Y.~J.}\ \bibnamefont {Guo}},\ }\href {\doibase 10.1063/1.3691191} {\bibfield  {journal} {\bibinfo  {journal} {Journal of Applied Physics}\ }\textbf {\bibinfo {volume} {111}},\ \bibinfo {pages} {053910} (\bibinfo {year} {2012})}\BibitemShut {NoStop}%
\bibitem [{\citenamefont {Pavlovskiy}\ \emph {et~al.}(2020)\citenamefont {Pavlovskiy}, \citenamefont {Gundareva}, \citenamefont {Volkov},\ and\ \citenamefont {Divin}}]{HTcJJde}%
  \BibitemOpen
  \bibfield  {author} {\bibinfo {author} {\bibfnamefont {V.~V.}\ \bibnamefont {Pavlovskiy}}, \bibinfo {author} {\bibfnamefont {I.~I.}\ \bibnamefont {Gundareva}}, \bibinfo {author} {\bibfnamefont {O.~Y.}\ \bibnamefont {Volkov}}, \ and\ \bibinfo {author} {\bibfnamefont {Y.~Y.}\ \bibnamefont {Divin}},\ }\href {\doibase 10.1063/1.5142400} {\bibfield  {journal} {\bibinfo  {journal} {Applied Physics Letters}\ }\textbf {\bibinfo {volume} {116}},\ \bibinfo {pages} {082601} (\bibinfo {year} {2020})}\BibitemShut {NoStop}%
\bibitem [{\citenamefont {Blonder}, \citenamefont {Tinkham},\ and\ \citenamefont {Klapwijk}(1982)}]{BTK}%
  \BibitemOpen
  \bibfield  {author} {\bibinfo {author} {\bibfnamefont {G.~E.}\ \bibnamefont {Blonder}}, \bibinfo {author} {\bibfnamefont {M.}~\bibnamefont {Tinkham}}, \ and\ \bibinfo {author} {\bibfnamefont {T.~M.}\ \bibnamefont {Klapwijk}},\ }\href {\doibase 10.1103/PhysRevB.25.4515} {\bibfield  {journal} {\bibinfo  {journal} {Physical Review B}\ }\textbf {\bibinfo {volume} {25}},\ \bibinfo {pages} {4515} (\bibinfo {year} {1982})}\BibitemShut {NoStop}%
\bibitem [{\citenamefont {Sullivan}\ \emph {et~al.}(1970)\citenamefont {Sullivan}, \citenamefont {Peterson}, \citenamefont {Kose},\ and\ \citenamefont {Zimmerman}}]{half-integ-theory}%
  \BibitemOpen
  \bibfield  {author} {\bibinfo {author} {\bibfnamefont {D.~B.}\ \bibnamefont {Sullivan}}, \bibinfo {author} {\bibfnamefont {R.~L.}\ \bibnamefont {Peterson}}, \bibinfo {author} {\bibfnamefont {V.~E.}\ \bibnamefont {Kose}}, \ and\ \bibinfo {author} {\bibfnamefont {J.~E.}\ \bibnamefont {Zimmerman}},\ }\href {\doibase 10.1063/1.1658554} {\bibfield  {journal} {\bibinfo  {journal} {Journal of Applied Physics}\ }\textbf {\bibinfo {volume} {41}},\ \bibinfo {pages} {4865} (\bibinfo {year} {1970})}\BibitemShut {NoStop}%
\bibitem [{\citenamefont {Golubov}, \citenamefont {Kupriyanov},\ and\ \citenamefont {Il’ichev}(1998)}]{hilgenkamp}%
  \BibitemOpen
  \bibfield  {author} {\bibinfo {author} {\bibfnamefont {A.~A.}\ \bibnamefont {Golubov}}, \bibinfo {author} {\bibfnamefont {M.~Y.}\ \bibnamefont {Kupriyanov}}, \ and\ \bibinfo {author} {\bibfnamefont {E.}~\bibnamefont {Il’ichev}},\ }\href {\doibase 10.1103/PhysRevLett.81.894} {\bibfield  {journal} {\bibinfo  {journal} {Physical Review Letters}\ }\textbf {\bibinfo {volume} {81}},\ \bibinfo {pages} {894} (\bibinfo {year} {1998})}\BibitemShut {NoStop}%
\bibitem [{\citenamefont {Murakami}\ \emph {et~al.}(1996)\citenamefont {Murakami}, \citenamefont {Akasako}, \citenamefont {Seto} \emph {et~al.}}]{point_contact}%
  \BibitemOpen
  \bibfield  {author} {\bibinfo {author} {\bibfnamefont {H.}~\bibnamefont {Murakami}}, \bibinfo {author} {\bibfnamefont {D.}~\bibnamefont {Akasako}}, \bibinfo {author} {\bibfnamefont {M.}~\bibnamefont {Seto}},  \emph {et~al.},\ }\href {\doibase 10.1007/BF02562787} {\bibfield  {journal} {\bibinfo  {journal} {Czech Journal of Physics}\ }\textbf {\bibinfo {volume} {46}},\ \bibinfo {pages} {1345} (\bibinfo {year} {1996})}\BibitemShut {NoStop}%
\bibitem [{\citenamefont {Du}\ \emph {et~al.}(2014)\citenamefont {Du}, \citenamefont {Lazar}, \citenamefont {Lam}, \citenamefont {Mitchell},\ and\ \citenamefont {Foley}}]{du2014fabrication}%
  \BibitemOpen
  \bibfield  {author} {\bibinfo {author} {\bibfnamefont {J.}~\bibnamefont {Du}}, \bibinfo {author} {\bibfnamefont {J.}~\bibnamefont {Lazar}}, \bibinfo {author} {\bibfnamefont {S.}~\bibnamefont {Lam}}, \bibinfo {author} {\bibfnamefont {E.}~\bibnamefont {Mitchell}}, \ and\ \bibinfo {author} {\bibfnamefont {C.}~\bibnamefont {Foley}},\ }\href@noop {} {\bibfield  {journal} {\bibinfo  {journal} {Superconductor Science and Technology}\ }\textbf {\bibinfo {volume} {27}},\ \bibinfo {pages} {095005} (\bibinfo {year} {2014})}\BibitemShut {NoStop}%
\bibitem [{\citenamefont {Cybart}, \citenamefont {Chen},\ and\ \citenamefont {Dynes}(2005)}]{cybart2005josephson}%
  \BibitemOpen
  \bibfield  {author} {\bibinfo {author} {\bibfnamefont {S.}~\bibnamefont {Cybart}}, \bibinfo {author} {\bibfnamefont {K.}~\bibnamefont {Chen}}, \ and\ \bibinfo {author} {\bibfnamefont {R.}~\bibnamefont {Dynes}},\ }\href@noop {} {\bibfield  {journal} {\bibinfo  {journal} {IEEE Trans. Appl. Supercon.}\ }\textbf {\bibinfo {volume} {15}},\ \bibinfo {pages} {241} (\bibinfo {year} {2005})}\BibitemShut {NoStop}%
\bibitem [{\citenamefont {Cou{\"e}do}\ \emph {et~al.}(2020)\citenamefont {Cou{\"e}do}, \citenamefont {Amari}, \citenamefont {Feuillet-Palma} \emph {et~al.}}]{franch_scientificreport}%
  \BibitemOpen
  \bibfield  {author} {\bibinfo {author} {\bibfnamefont {F.}~\bibnamefont {Cou{\"e}do}}, \bibinfo {author} {\bibfnamefont {P.}~\bibnamefont {Amari}}, \bibinfo {author} {\bibfnamefont {C.}~\bibnamefont {Feuillet-Palma}},  \emph {et~al.},\ }\href@noop {} {\bibfield  {journal} {\bibinfo  {journal} {Scientific Reports}\ }\textbf {\bibinfo {volume} {10}},\ \bibinfo {pages} {10256} (\bibinfo {year} {2020})}\BibitemShut {NoStop}%
\bibitem [{\citenamefont {Chen}\ \emph {et~al.}(2022)\citenamefont {Chen}, \citenamefont {Li}, \citenamefont {Zhu}, \citenamefont {Xu}, \citenamefont {Xu}, \citenamefont {Yin}, \citenamefont {Cai}, \citenamefont {Wang}, \citenamefont {Lu}, \citenamefont {Zhang},\ and\ \citenamefont {Ma}}]{Chinease-chen2022high}%
  \BibitemOpen
  \bibfield  {author} {\bibinfo {author} {\bibfnamefont {Z.}~\bibnamefont {Chen}}, \bibinfo {author} {\bibfnamefont {Y.}~\bibnamefont {Li}}, \bibinfo {author} {\bibfnamefont {R.}~\bibnamefont {Zhu}}, \bibinfo {author} {\bibfnamefont {J.}~\bibnamefont {Xu}}, \bibinfo {author} {\bibfnamefont {T.}~\bibnamefont {Xu}}, \bibinfo {author} {\bibfnamefont {D.}~\bibnamefont {Yin}}, \bibinfo {author} {\bibfnamefont {X.}~\bibnamefont {Cai}}, \bibinfo {author} {\bibfnamefont {Y.}~\bibnamefont {Wang}}, \bibinfo {author} {\bibfnamefont {J.}~\bibnamefont {Lu}}, \bibinfo {author} {\bibfnamefont {Y.}~\bibnamefont {Zhang}}, \ and\ \bibinfo {author} {\bibfnamefont {P.}~\bibnamefont {Ma}},\ }\href {\doibase 10.1088/0256-307X/39/7/077402} {\bibfield  {journal} {\bibinfo  {journal} {Chinese Physics Letters}\ }\textbf {\bibinfo {volume} {39}},\ \bibinfo {pages} {077402} (\bibinfo {year} {2022})}\BibitemShut {NoStop}%
\bibitem [{\citenamefont {Cybart}\ \emph {et~al.}(2015)\citenamefont {Cybart}, \citenamefont {Cho}, \citenamefont {Wong}, \citenamefont {Wehlin}, \citenamefont {Ma}, \citenamefont {Huynh},\ and\ \citenamefont {Dynes}}]{cybart2015nano}%
  \BibitemOpen
  \bibfield  {author} {\bibinfo {author} {\bibfnamefont {S.~A.}\ \bibnamefont {Cybart}}, \bibinfo {author} {\bibfnamefont {E.}~\bibnamefont {Cho}}, \bibinfo {author} {\bibfnamefont {T.}~\bibnamefont {Wong}}, \bibinfo {author} {\bibfnamefont {B.~H.}\ \bibnamefont {Wehlin}}, \bibinfo {author} {\bibfnamefont {M.~K.}\ \bibnamefont {Ma}}, \bibinfo {author} {\bibfnamefont {C.}~\bibnamefont {Huynh}}, \ and\ \bibinfo {author} {\bibfnamefont {R.}~\bibnamefont {Dynes}},\ }\href@noop {} {\bibfield  {journal} {\bibinfo  {journal} {Nat. Nanotechnol.}\ }\textbf {\bibinfo {volume} {10}},\ \bibinfo {pages} {598} (\bibinfo {year} {2015})}\BibitemShut {NoStop}%
\bibitem [{\citenamefont {Cho}\ \emph {et~al.}(2018)\citenamefont {Cho}, \citenamefont {Zhou}, \citenamefont {Cho},\ and\ \citenamefont {Cybart}}]{cho2018superconducting}%
  \BibitemOpen
  \bibfield  {author} {\bibinfo {author} {\bibfnamefont {E.~Y.}\ \bibnamefont {Cho}}, \bibinfo {author} {\bibfnamefont {Y.~W.}\ \bibnamefont {Zhou}}, \bibinfo {author} {\bibfnamefont {J.~Y.}\ \bibnamefont {Cho}}, \ and\ \bibinfo {author} {\bibfnamefont {S.~A.}\ \bibnamefont {Cybart}},\ }\href@noop {} {\bibfield  {journal} {\bibinfo  {journal} {Appl. Phys. Lett.}\ }\textbf {\bibinfo {volume} {113}},\ \bibinfo {pages} {022604} (\bibinfo {year} {2018})}\BibitemShut {NoStop}%
\bibitem [{\citenamefont {LeFebvre}, \citenamefont {Parachikunnumal},\ and\ \citenamefont {Cybart}(2024)}]{shaznjay}%
  \BibitemOpen
  \bibfield  {author} {\bibinfo {author} {\bibfnamefont {J.~C.}\ \bibnamefont {LeFebvre}}, \bibinfo {author} {\bibfnamefont {A.}~\bibnamefont {Parachikunnumal}}, \ and\ \bibinfo {author} {\bibfnamefont {S.~A.}\ \bibnamefont {Cybart}},\ }\href {\doibase 10.1116/6.0004020} {\bibfield  {journal} {\bibinfo  {journal} {Journal of Vacuum Science \& Technology B}\ }\textbf {\bibinfo {volume} {42}},\ \bibinfo {pages} {063204} (\bibinfo {year} {2024})}\BibitemShut {NoStop}%
\bibitem [{\citenamefont {Cho}\ \emph {et~al.}(2015{\natexlab{a}})\citenamefont {Cho}, \citenamefont {Ma}, \citenamefont {Huynh}, \citenamefont {Pratt}, \citenamefont {Paulson}, \citenamefont {Glyantsev}, \citenamefont {Dynes},\ and\ \citenamefont {Cybart}}]{ethan_squid}%
  \BibitemOpen
  \bibfield  {author} {\bibinfo {author} {\bibfnamefont {E.~Y.}\ \bibnamefont {Cho}}, \bibinfo {author} {\bibfnamefont {M.~K.}\ \bibnamefont {Ma}}, \bibinfo {author} {\bibfnamefont {C.}~\bibnamefont {Huynh}}, \bibinfo {author} {\bibfnamefont {K.}~\bibnamefont {Pratt}}, \bibinfo {author} {\bibfnamefont {D.~N.}\ \bibnamefont {Paulson}}, \bibinfo {author} {\bibfnamefont {V.~N.}\ \bibnamefont {Glyantsev}}, \bibinfo {author} {\bibfnamefont {R.~C.}\ \bibnamefont {Dynes}}, \ and\ \bibinfo {author} {\bibfnamefont {S.~A.}\ \bibnamefont {Cybart}},\ }\href@noop {} {\bibfield  {journal} {\bibinfo  {journal} {Appl. Phys. Lett.}\ }\textbf {\bibinfo {volume} {106}},\ \bibinfo {pages} {252601} (\bibinfo {year} {2015}{\natexlab{a}})}\BibitemShut {NoStop}%
\bibitem [{\citenamefont {Cybart}\ \emph {et~al.}(2013)\citenamefont {Cybart}, \citenamefont {Roediger}, \citenamefont {Chen}, \citenamefont {Parker}, \citenamefont {Cho}, \citenamefont {Wong},\ and\ \citenamefont {Dynes}}]{cybartstability}%
  \BibitemOpen
  \bibfield  {author} {\bibinfo {author} {\bibfnamefont {S.~A.}\ \bibnamefont {Cybart}}, \bibinfo {author} {\bibfnamefont {P.}~\bibnamefont {Roediger}}, \bibinfo {author} {\bibfnamefont {K.}~\bibnamefont {Chen}}, \bibinfo {author} {\bibfnamefont {J.~M.}\ \bibnamefont {Parker}}, \bibinfo {author} {\bibfnamefont {E.~Y.}\ \bibnamefont {Cho}}, \bibinfo {author} {\bibfnamefont {T.~J.}\ \bibnamefont {Wong}}, \ and\ \bibinfo {author} {\bibfnamefont {R.~C.}\ \bibnamefont {Dynes}},\ }\href {\doibase 10.1109/TASC.2012.2227646} {\bibfield  {journal} {\bibinfo  {journal} {IEEE Transactions on Applied Superconductivity}\ }\textbf {\bibinfo {volume} {23}},\ \bibinfo {pages} {1100103} (\bibinfo {year} {2013})}\BibitemShut {NoStop}%
\bibitem [{\citenamefont {M{\"u}ller}\ \emph {et~al.}(2019)\citenamefont {M{\"u}ller}, \citenamefont {Karrer}, \citenamefont {Limberger}, \citenamefont {Becker}, \citenamefont {Schr{\"o}ppel}, \citenamefont {Burkhardt}, \citenamefont {Kleiner}, \citenamefont {Goldobin},\ and\ \citenamefont {Koelle}}]{germany-muller2019josephson}%
  \BibitemOpen
  \bibfield  {author} {\bibinfo {author} {\bibfnamefont {B.}~\bibnamefont {M{\"u}ller}}, \bibinfo {author} {\bibfnamefont {M.}~\bibnamefont {Karrer}}, \bibinfo {author} {\bibfnamefont {F.}~\bibnamefont {Limberger}}, \bibinfo {author} {\bibfnamefont {M.}~\bibnamefont {Becker}}, \bibinfo {author} {\bibfnamefont {B.}~\bibnamefont {Schr{\"o}ppel}}, \bibinfo {author} {\bibfnamefont {C.~J.}\ \bibnamefont {Burkhardt}}, \bibinfo {author} {\bibfnamefont {R.}~\bibnamefont {Kleiner}}, \bibinfo {author} {\bibfnamefont {E.}~\bibnamefont {Goldobin}}, \ and\ \bibinfo {author} {\bibfnamefont {D.}~\bibnamefont {Koelle}},\ }\href {\doibase 10.1103/PhysRevApplied.11.044082} {\bibfield  {journal} {\bibinfo  {journal} {Physical Review Applied}\ }\textbf {\bibinfo {volume} {11}},\ \bibinfo {pages} {044082} (\bibinfo {year} {2019})}\BibitemShut {NoStop}%
\bibitem [{\citenamefont {Cho}\ \emph {et~al.}(2015{\natexlab{b}})\citenamefont {Cho}, \citenamefont {Ma}, \citenamefont {Huynh}, \citenamefont {Pratt}, \citenamefont {Paulson}, \citenamefont {Glyantsev}, \citenamefont {Dynes},\ and\ \citenamefont {Cybart}}]{cho2015yba2cu3o7}%
  \BibitemOpen
  \bibfield  {author} {\bibinfo {author} {\bibfnamefont {E.}~\bibnamefont {Cho}}, \bibinfo {author} {\bibfnamefont {M.}~\bibnamefont {Ma}}, \bibinfo {author} {\bibfnamefont {C.}~\bibnamefont {Huynh}}, \bibinfo {author} {\bibfnamefont {K.}~\bibnamefont {Pratt}}, \bibinfo {author} {\bibfnamefont {D.}~\bibnamefont {Paulson}}, \bibinfo {author} {\bibfnamefont {V.}~\bibnamefont {Glyantsev}}, \bibinfo {author} {\bibfnamefont {R.}~\bibnamefont {Dynes}}, \ and\ \bibinfo {author} {\bibfnamefont {S.~A.}\ \bibnamefont {Cybart}},\ }\href@noop {} {\bibfield  {journal} {\bibinfo  {journal} {Appl. Phys. Lett.}\ }\textbf {\bibinfo {volume} {106}},\ \bibinfo {pages} {252601} (\bibinfo {year} {2015}{\natexlab{b}})}\BibitemShut {NoStop}%
\bibitem [{\citenamefont {LeFebvre}\ and\ \citenamefont {Cybart}(2021)}]{jay2021LargeScaleHIBL}%
  \BibitemOpen
  \bibfield  {author} {\bibinfo {author} {\bibfnamefont {J.~C.}\ \bibnamefont {LeFebvre}}\ and\ \bibinfo {author} {\bibfnamefont {S.~A.}\ \bibnamefont {Cybart}},\ }\href {\doibase 10.1109/TASC.2021.3057324} {\bibfield  {journal} {\bibinfo  {journal} {IEEE Transactions on Applied Superconductivity}\ }\textbf {\bibinfo {volume} {31}},\ \bibinfo {pages} {1} (\bibinfo {year} {2021})}\BibitemShut {NoStop}%
\bibitem [{\citenamefont {Cybart}\ \emph {et~al.}(2009)\citenamefont {Cybart}, \citenamefont {Anton}, \citenamefont {Wu}, \citenamefont {Clarke},\ and\ \citenamefont {Dynes}}]{Cybart2009nanolette}%
  \BibitemOpen
  \bibfield  {author} {\bibinfo {author} {\bibfnamefont {S.~A.}\ \bibnamefont {Cybart}}, \bibinfo {author} {\bibfnamefont {S.~M.}\ \bibnamefont {Anton}}, \bibinfo {author} {\bibfnamefont {S.~M.}\ \bibnamefont {Wu}}, \bibinfo {author} {\bibfnamefont {J.}~\bibnamefont {Clarke}}, \ and\ \bibinfo {author} {\bibfnamefont {R.~C.}\ \bibnamefont {Dynes}},\ }\href {\doibase 10.1021/nl901785j} {\bibfield  {journal} {\bibinfo  {journal} {Nano Letters}\ }\textbf {\bibinfo {volume} {9}},\ \bibinfo {pages} {3535–3539} (\bibinfo {year} {2009})}\BibitemShut {NoStop}%
\bibitem [{\citenamefont {LeFebvre}\ \emph {et~al.}(2022)\citenamefont {LeFebvre}, \citenamefont {Cho}, \citenamefont {Li}, \citenamefont {Cai},\ and\ \citenamefont {Cybart}}]{longjay}%
  \BibitemOpen
  \bibfield  {author} {\bibinfo {author} {\bibfnamefont {J.~C.}\ \bibnamefont {LeFebvre}}, \bibinfo {author} {\bibfnamefont {E.~Y.}\ \bibnamefont {Cho}}, \bibinfo {author} {\bibfnamefont {H.}~\bibnamefont {Li}}, \bibinfo {author} {\bibfnamefont {H.}~\bibnamefont {Cai}}, \ and\ \bibinfo {author} {\bibfnamefont {S.~A.}\ \bibnamefont {Cybart}},\ }\href {\doibase 10.1063/5.0087611} {\bibfield  {journal} {\bibinfo  {journal} {Journal of Applied Physics}\ }\textbf {\bibinfo {volume} {131}},\ \bibinfo {pages} {163902} (\bibinfo {year} {2022})}\BibitemShut {NoStop}%
\bibitem [{\citenamefont {Cho}\ \emph {et~al.}(2019)\citenamefont {Cho}, \citenamefont {Zhou}, \citenamefont {Khapaev},\ and\ \citenamefont {Cybart}}]{Cho20192DHighTcSQUIDArrays}%
  \BibitemOpen
  \bibfield  {author} {\bibinfo {author} {\bibfnamefont {E.~Y.}\ \bibnamefont {Cho}}, \bibinfo {author} {\bibfnamefont {Y.~W.}\ \bibnamefont {Zhou}}, \bibinfo {author} {\bibfnamefont {M.~M.}\ \bibnamefont {Khapaev}}, \ and\ \bibinfo {author} {\bibfnamefont {S.~A.}\ \bibnamefont {Cybart}},\ }\href {\doibase 10.1109/TASC.2019.2904481} {\bibfield  {journal} {\bibinfo  {journal} {IEEE Transactions on Applied Superconductivity}\ }\textbf {\bibinfo {volume} {29}},\ \bibinfo {pages} {1} (\bibinfo {year} {2019})}\BibitemShut {NoStop}%
\bibitem [{\citenamefont {Li}\ \emph {et~al.}(2024)\citenamefont {Li}, \citenamefont {Cai}, \citenamefont {Sarkar}, \citenamefont {LeFebvre}, \citenamefont {Cho},\ and\ \citenamefont {Cybart}}]{haosquidarray-APL}%
  \BibitemOpen
  \bibfield  {author} {\bibinfo {author} {\bibfnamefont {H.}~\bibnamefont {Li}}, \bibinfo {author} {\bibfnamefont {H.}~\bibnamefont {Cai}}, \bibinfo {author} {\bibfnamefont {N.}~\bibnamefont {Sarkar}}, \bibinfo {author} {\bibfnamefont {J.~C.}\ \bibnamefont {LeFebvre}}, \bibinfo {author} {\bibfnamefont {E.~Y.}\ \bibnamefont {Cho}}, \ and\ \bibinfo {author} {\bibfnamefont {S.~A.}\ \bibnamefont {Cybart}},\ }\href {\doibase 10.1063/5.0206821} {\bibfield  {journal} {\bibinfo  {journal} {Applied Physics Letters}\ }\textbf {\bibinfo {volume} {124}},\ \bibinfo {pages} {192603} (\bibinfo {year} {2024})}\BibitemShut {NoStop}%
\bibitem [{\citenamefont {Cai}\ \emph {et~al.}(2024)\citenamefont {Cai}, \citenamefont {LeFebvre}, \citenamefont {Li}, \citenamefont {Cho}, \citenamefont {Yoshikawa},\ and\ \citenamefont {Cybart}}]{hanqfp}%
  \BibitemOpen
  \bibfield  {author} {\bibinfo {author} {\bibfnamefont {H.}~\bibnamefont {Cai}}, \bibinfo {author} {\bibfnamefont {J.~C.}\ \bibnamefont {LeFebvre}}, \bibinfo {author} {\bibfnamefont {H.}~\bibnamefont {Li}}, \bibinfo {author} {\bibfnamefont {E.~Y.}\ \bibnamefont {Cho}}, \bibinfo {author} {\bibfnamefont {N.}~\bibnamefont {Yoshikawa}}, \ and\ \bibinfo {author} {\bibfnamefont {S.~A.}\ \bibnamefont {Cybart}},\ }\href {\doibase 10.1063/5.0206445} {\bibfield  {journal} {\bibinfo  {journal} {Applied Physics Letters}\ }\textbf {\bibinfo {volume} {124}},\ \bibinfo {pages} {212601} (\bibinfo {year} {2024})}\BibitemShut {NoStop}%
\bibitem [{\citenamefont {Cai}\ \emph {et~al.}(2021)\citenamefont {Cai}, \citenamefont {Li}, \citenamefont {Cho}, \citenamefont {LeFebvre},\ and\ \citenamefont {Cybart}}]{hna_SFQ}%
  \BibitemOpen
  \bibfield  {author} {\bibinfo {author} {\bibfnamefont {H.}~\bibnamefont {Cai}}, \bibinfo {author} {\bibfnamefont {H.}~\bibnamefont {Li}}, \bibinfo {author} {\bibfnamefont {E.~Y.}\ \bibnamefont {Cho}}, \bibinfo {author} {\bibfnamefont {J.~C.}\ \bibnamefont {LeFebvre}}, \ and\ \bibinfo {author} {\bibfnamefont {S.~A.}\ \bibnamefont {Cybart}},\ }\href {\doibase 10.1109/TASC.2021.3058088} {\bibfield  {journal} {\bibinfo  {journal} {IEEE Transactions on Applied Superconductivity}\ }\textbf {\bibinfo {volume} {31}},\ \bibinfo {pages} {1} (\bibinfo {year} {2021})}\BibitemShut {NoStop}%
\bibitem [{\citenamefont {Wu}\ \emph {et~al.}(2024)\citenamefont {Wu}, \citenamefont {Hack}, \citenamefont {Wurster}, \citenamefont {Koch}, \citenamefont {Kleiner}, \citenamefont {Koelle}, \citenamefont {Mannhart},\ and\ \citenamefont {Harbola}}]{VectorSubstrateJJ}%
  \BibitemOpen
  \bibfield  {author} {\bibinfo {author} {\bibfnamefont {Y.-J.}\ \bibnamefont {Wu}}, \bibinfo {author} {\bibfnamefont {M.}~\bibnamefont {Hack}}, \bibinfo {author} {\bibfnamefont {K.}~\bibnamefont {Wurster}}, \bibinfo {author} {\bibfnamefont {S.}~\bibnamefont {Koch}}, \bibinfo {author} {\bibfnamefont {R.}~\bibnamefont {Kleiner}}, \bibinfo {author} {\bibfnamefont {D.}~\bibnamefont {Koelle}}, \bibinfo {author} {\bibfnamefont {J.}~\bibnamefont {Mannhart}}, \ and\ \bibinfo {author} {\bibfnamefont {V.}~\bibnamefont {Harbola}},\ }\href {\doibase 10.1063/5.0217861} {\bibfield  {journal} {\bibinfo  {journal} {Applied Physics Letters}\ }\textbf {\bibinfo {volume} {125}},\ \bibinfo {pages} {032601} (\bibinfo {year} {2024})}\BibitemShut {NoStop}%
\bibitem [{\citenamefont {Schmid}\ \emph {et~al.}(2025)\citenamefont {Schmid}, \citenamefont {Jozani}, \citenamefont {Kleiner}, \citenamefont {Koelle},\ and\ \citenamefont {Goldobin}}]{YBCOJosephsonDiode_dieter}%
  \BibitemOpen
  \bibfield  {author} {\bibinfo {author} {\bibfnamefont {C.}~\bibnamefont {Schmid}}, \bibinfo {author} {\bibfnamefont {A.}~\bibnamefont {Jozani}}, \bibinfo {author} {\bibfnamefont {R.}~\bibnamefont {Kleiner}}, \bibinfo {author} {\bibfnamefont {D.}~\bibnamefont {Koelle}}, \ and\ \bibinfo {author} {\bibfnamefont {E.}~\bibnamefont {Goldobin}},\ }\href {\doibase 10.1103/PhysRevApplied.24.014041} {\bibfield  {journal} {\bibinfo  {journal} {Physical Review Applied}\ }\textbf {\bibinfo {volume} {24}},\ \bibinfo {pages} {014041} (\bibinfo {year} {2025})}\BibitemShut {NoStop}%
\bibitem [{\citenamefont {Pröpper}\ \emph {et~al.}(2025)\citenamefont {Pröpper}, \citenamefont {Hanisch}, \citenamefont {Schmid}, \citenamefont {Neumann}, \citenamefont {Ritter}, \citenamefont {Tucholke}, \citenamefont {Goldobin}, \citenamefont {Koelle}, \citenamefont {Kleiner}, \citenamefont {Schilling},\ and\ \citenamefont {Hampel}}]{Max_HeFIBJJA}%
  \BibitemOpen
  \bibfield  {author} {\bibinfo {author} {\bibfnamefont {M.}~\bibnamefont {Pröpper}}, \bibinfo {author} {\bibfnamefont {D.}~\bibnamefont {Hanisch}}, \bibinfo {author} {\bibfnamefont {C.}~\bibnamefont {Schmid}}, \bibinfo {author} {\bibfnamefont {M.}~\bibnamefont {Neumann}}, \bibinfo {author} {\bibfnamefont {P.~J.}\ \bibnamefont {Ritter}}, \bibinfo {author} {\bibfnamefont {M.-A.}\ \bibnamefont {Tucholke}}, \bibinfo {author} {\bibfnamefont {E.}~\bibnamefont {Goldobin}}, \bibinfo {author} {\bibfnamefont {D.}~\bibnamefont {Koelle}}, \bibinfo {author} {\bibfnamefont {R.}~\bibnamefont {Kleiner}}, \bibinfo {author} {\bibfnamefont {M.}~\bibnamefont {Schilling}}, \ and\ \bibinfo {author} {\bibfnamefont {B.}~\bibnamefont {Hampel}},\ }\href {\doibase 10.1109/TASC.2024.3516741} {\bibfield  {journal} {\bibinfo  {journal} {IEEE Transactions on Applied Superconductivity}\ }\textbf {\bibinfo {volume} {35}},\ \bibinfo {pages} {1} (\bibinfo {year} {2025})}\BibitemShut {NoStop}%
\bibitem [{\citenamefont {Cortez}\ \emph {et~al.}(2019)\citenamefont {Cortez}, \citenamefont {Cho}, \citenamefont {Li}, \citenamefont {Cunnane}, \citenamefont {Karasik},\ and\ \citenamefont {Cybart}}]{cortez2019tuning}%
  \BibitemOpen
  \bibfield  {author} {\bibinfo {author} {\bibfnamefont {A.~T.}\ \bibnamefont {Cortez}}, \bibinfo {author} {\bibfnamefont {E.~Y.}\ \bibnamefont {Cho}}, \bibinfo {author} {\bibfnamefont {H.}~\bibnamefont {Li}}, \bibinfo {author} {\bibfnamefont {D.}~\bibnamefont {Cunnane}}, \bibinfo {author} {\bibfnamefont {B.}~\bibnamefont {Karasik}}, \ and\ \bibinfo {author} {\bibfnamefont {S.~A.}\ \bibnamefont {Cybart}},\ }\href {\doibase 10.1109/TASC.2019.2905166} {\bibfield  {journal} {\bibinfo  {journal} {IEEE Transactions on Applied Superconductivity}\ }\textbf {\bibinfo {volume} {29}},\ \bibinfo {pages} {1} (\bibinfo {year} {2019})}\BibitemShut {NoStop}%
\bibitem [{\citenamefont {Wang}\ \emph {et~al.}(2019{\natexlab{a}})\citenamefont {Wang}, \citenamefont {Semerad}, \citenamefont {McCoy}, \citenamefont {Cai}, \citenamefont {LeFebvre}, \citenamefont {Grezdo}, \citenamefont {Cho}, \citenamefont {Li},\ and\ \citenamefont {Cybart}}]{YT_Multilayers}%
  \BibitemOpen
  \bibfield  {author} {\bibinfo {author} {\bibfnamefont {Y.-T.}\ \bibnamefont {Wang}}, \bibinfo {author} {\bibfnamefont {R.}~\bibnamefont {Semerad}}, \bibinfo {author} {\bibfnamefont {S.~J.}\ \bibnamefont {McCoy}}, \bibinfo {author} {\bibfnamefont {H.}~\bibnamefont {Cai}}, \bibinfo {author} {\bibfnamefont {J.}~\bibnamefont {LeFebvre}}, \bibinfo {author} {\bibfnamefont {H.}~\bibnamefont {Grezdo}}, \bibinfo {author} {\bibfnamefont {E.~Y.}\ \bibnamefont {Cho}}, \bibinfo {author} {\bibfnamefont {H.}~\bibnamefont {Li}}, \ and\ \bibinfo {author} {\bibfnamefont {S.~A.}\ \bibnamefont {Cybart}},\ }\href {\doibase 10.1109/TASC.2019.2898778} {\bibfield  {journal} {\bibinfo  {journal} {IEEE Transactions on Applied Superconductivity}\ }\textbf {\bibinfo {volume} {29}},\ \bibinfo {pages} {1} (\bibinfo {year} {2019}{\natexlab{a}})}\BibitemShut {NoStop}%
\bibitem [{\citenamefont {Wang}\ \emph {et~al.}(2019{\natexlab{b}})\citenamefont {Wang}, \citenamefont {Cho}, \citenamefont {Li},\ and\ \citenamefont {Cybart}}]{yan-tin_jj_width}%
  \BibitemOpen
  \bibfield  {author} {\bibinfo {author} {\bibfnamefont {Y.-T.}\ \bibnamefont {Wang}}, \bibinfo {author} {\bibfnamefont {E.~Y.}\ \bibnamefont {Cho}}, \bibinfo {author} {\bibfnamefont {H.}~\bibnamefont {Li}}, \ and\ \bibinfo {author} {\bibfnamefont {S.~A.}\ \bibnamefont {Cybart}},\ }in\ \href {\doibase 10.1109/ISEC46533.2019.8990944} {\emph {\bibinfo {booktitle} {2019 IEEE International Superconductive Electronics Conference (ISEC)}}}\ (\bibinfo  {publisher} {IEEE},\ \bibinfo {year} {2019})\ pp.\ \bibinfo {pages} {1--3}\BibitemShut {NoStop}%
\bibitem [{\citenamefont {de~Gennes}(1964)}]{deGennes1964}%
  \BibitemOpen
  \bibfield  {author} {\bibinfo {author} {\bibfnamefont {P.~G.}\ \bibnamefont {de~Gennes}},\ }\href {\doibase 10.1103/RevModPhys.36.225} {\bibfield  {journal} {\bibinfo  {journal} {Reviews of Modern Physics}\ }\textbf {\bibinfo {volume} {36}},\ \bibinfo {pages} {225} (\bibinfo {year} {1964})}\BibitemShut {NoStop}%
\bibitem [{\citenamefont {Tolpygo}\ and\ \citenamefont {Gurvitch}(1996)}]{Tolpygo1996CriticalCurrents}%
  \BibitemOpen
  \bibfield  {author} {\bibinfo {author} {\bibfnamefont {S.~K.}\ \bibnamefont {Tolpygo}}\ and\ \bibinfo {author} {\bibfnamefont {M.}~\bibnamefont {Gurvitch}},\ }\href {\doibase 10.1063/1.117568} {\bibfield  {journal} {\bibinfo  {journal} {Applied Physics Letters}\ }\textbf {\bibinfo {volume} {69}},\ \bibinfo {pages} {3914} (\bibinfo {year} {1996})}\BibitemShut {NoStop}%
\bibitem [{\citenamefont {Kautz}(1995)}]{kautz1995shapiro}%
  \BibitemOpen
  \bibfield  {author} {\bibinfo {author} {\bibfnamefont {R.~L.}\ \bibnamefont {Kautz}},\ }\href {\doibase 10.1063/1.359644} {\bibfield  {journal} {\bibinfo  {journal} {Journal of Applied Physics}\ }\textbf {\bibinfo {volume} {78}},\ \bibinfo {pages} {5811} (\bibinfo {year} {1995})}\BibitemShut {NoStop}%
\bibitem [{\citenamefont {McCumber}(1968)}]{mccumber1968effect}%
  \BibitemOpen
  \bibfield  {author} {\bibinfo {author} {\bibfnamefont {D.}~\bibnamefont {McCumber}},\ }\href@noop {} {\bibfield  {journal} {\bibinfo  {journal} {Journal of Applied Physics}\ }\textbf {\bibinfo {volume} {39}},\ \bibinfo {pages} {3113} (\bibinfo {year} {1968})}\BibitemShut {NoStop}%
\end{thebibliography}

\begin{thebibliography}{37}%
\makeatletter
\providecommand \@ifxundefined [1]{%
 \@ifx{#1\undefined}
}%
\providecommand \@ifnum [1]{%
 \ifnum #1\expandafter \@firstoftwo
 \else \expandafter \@secondoftwo
 \fi
}%
\providecommand \@ifx [1]{%
 \ifx #1\expandafter \@firstoftwo
 \else \expandafter \@secondoftwo
 \fi
}%
\providecommand \natexlab [1]{#1}%
\providecommand \enquote  [1]{``#1''}%
\providecommand \bibnamefont  [1]{#1}%
\providecommand \bibfnamefont [1]{#1}%
\providecommand \citenamefont [1]{#1}%
\providecommand \href@noop [0]{\@secondoftwo}%
\providecommand \href [0]{\begingroup \@sanitize@url \@href}%
\providecommand \@href[1]{\@@startlink{#1}\@@href}%
\providecommand \@@href[1]{\endgroup#1\@@endlink}%
\providecommand \@sanitize@url [0]{\catcode `\\12\catcode `\$12\catcode `\&12\catcode `\#12\catcode `\^12\catcode `\_12\catcode `\%12\relax}%
\providecommand \@@startlink[1]{}%
\providecommand \@@endlink[0]{}%
\providecommand \url  [0]{\begingroup\@sanitize@url \@url }%
\providecommand \@url [1]{\endgroup\@href {#1}{\urlprefix }}%
\providecommand \urlprefix  [0]{URL }%
\providecommand \Eprint [0]{\href }%
\providecommand \doibase [0]{http://dx.doi.org/}%
\providecommand \selectlanguage [0]{\@gobble}%
\providecommand \bibinfo  [0]{\@secondoftwo}%
\providecommand \bibfield  [0]{\@secondoftwo}%
\providecommand \translation [1]{[#1]}%
\providecommand \BibitemOpen [0]{}%
\providecommand \bibitemStop [0]{}%
\providecommand \bibitemNoStop [0]{.\EOS\space}%
\providecommand \EOS [0]{\spacefactor3000\relax}%
\providecommand \BibitemShut  [1]{\csname bibitem#1\endcsname}%
\let\auto@bib@innerbib\@empty
%</preamble>

\bibitem [{\citenamefont {Marchegiani}, \citenamefont {Braggio},\ and\ \citenamefont {Giazotto}(2020)}]{aravind_sug}%
  \BibitemOpen
  \bibfield  {author} {\bibinfo {author} {\bibfnamefont {G.}~\bibnamefont {Marchegiani}}, \bibinfo {author} {\bibfnamefont {A.}~\bibnamefont {Braggio}}, \ and\ \bibinfo {author} {\bibfnamefont {F.}~\bibnamefont {Giazotto}},\ }\href {\doibase 10.1063/5.0029984} {\bibfield  {journal} {\bibinfo  {journal} {Applied Physics Letters}\ }\textbf {\bibinfo {volume} {117}},\ \bibinfo {pages} {212601} (\bibinfo {year} {2020})}\BibitemShut {NoStop}%
\end{thebibliography}
%merlin.mbs aipnum4-1.bst 2010-07-25 4.21a (PWD, AO, DPC) hacked
%Control: key (0)
%Control: author (8) initials jnrlst
%Control: editor formatted (1) identically to author
%Control: production of article title (-1) disabled
%Control: page (0) single
%Control: year (1) truncated
%Control: production of eprint (0) enabled
%

%\end{document}

\end{document}